\documentclass[12pt,titlepage]{article}
\usepackage{geometry}
\usepackage{graphics}
\usepackage[pdftex] {graphicx}
\usepackage[usenames,dvipsnames]{color}
\usepackage[labelfont=bf, labelsep=space]{caption}
\usepackage{anysize}
\usepackage{amsmath}
\usepackage{bigstrut}
\usepackage[sort]{natbib}
\usepackage{colortbl}
\usepackage{color}
\usepackage{array}
\usepackage[table]{xcolor}
\usepackage{subcaption}
\usepackage{authblk}
\usepackage{setspace}
\usepackage{mathtools}
\definecolor{light-gray}{gray}{0.75}
\setcitestyle{aysep={},yysep={;}}

\usepackage{newfloat}
\DeclareFloatingEnvironment[name={Figure S}]{supfigure}

\begin{document}
\title{Probabilistic Population Projections for Countries with Generalized HIV/AIDS  Epidemics}

\author[1,5]{David J. Sharrow}
\author[2]{Jessica Godwin}
\author[2]{Yanjun He}
\author[3,4]{Samuel J. Clark}
\author[2,3]{Adrian E. Raftery}

\affil[1]{Center for Statistics and the Social Sciences, University of Washington}
\affil[2]{Department of Statistics, University of Washington}
\affil[3]{Department of Sociology, University of Washington}
\affil[4]{MRC/Wits Rural Public Health and Health Transitions Research Unit (Agincourt), School of Public Health, Faculty of Health Sciences, University of the Witwatersrand}
\affil[5] {\color{black} Corresponding author: dsharrow@uw.edu}

\maketitle
\thispagestyle{empty}
\newpage
\thispagestyle{empty}
\begin{abstract}
The United Nations (UN) issued official probabilistic population projections
for all countries to 2100 in July 2015.
This was done by simulating future levels of total fertility 
and life expectancy from Bayesian hierarchical models, and
combining the results using a standard cohort-component projection method.
The 40 countries with generalized HIV/AIDS epidemics were treated
differently from others, in that the projections used
the highly multistate Spectrum/EPP model, a complex 15-compartment model
that was designed for short-term projections of quantities relevant 
to policy for the epidemic. Here we propose a simpler approach that is more 
compatible with the existing UN probabilistic projection methodology for
other countries. Changes in life expectancy are projected probabilistically
using a simple time series regression model on current life expectancy, 
HIV prevalence and ART coverage. 
These are then converted to age- and sex-specific mortality rates
using a new family of model life tables designed for countries with
HIV/AIDS epidemics that reproduces the characteristic hump in 
middle adult mortality. These are then input to the standard
cohort-component method, as for other countries. 
The method performed well in an out-of-sample cross-validation 
experiment. It gives similar population projections to Spectrum/EPP in the short run,
while being simpler and avoiding multistate modeling. \\

\end{abstract}

\bibliographystyle{apa}

\clearpage
\setcounter{page}{1} 
\section{Introduction}
\linespread{1.5}

Population projections are used by governments at all levels, 
international organizations,  social and health researchers, and by the private sector 
for policy planning, monitoring development
goals and as inputs to economic and environmental models. 
The United Nations (UN) issues population projections for all the countries
of the world by age and sex to 2100, updated every two years, in the
{\it World Population Prospects} (WPP). It is the only organization to do so,
and its projections are the de facto standard at the global level
\citep{LutzSamir2010}.

There has long been great interest in probabilistic population projections, to
quantify uncertainty in projections and the risk of
adverse demographic events. However, the dominant population projection
methods are deterministic, and uncertainty has typically been conveyed
by variant scenario projections (such as High and Low projections) using 
different assumptions about future rates. This approach has 
been criticized as lacking validity because it has no probabilistic basis,
and leads to possible contradictions \citep{LeeTuljapurkar1994,NRC2000}.

In a major step forward, the UN issued official probabilistic population
projections for all countries in July 2015,
available at http://esa.un.org/unpd/ppp.
These projections were developed using the methodology of \citet{raftery2012}. 
They take account of uncertainty about
future levels of total fertility and life expectancy, using the
Bayesian hierarchical models of \citet{alkema2011} for fertility
and \citet{raftery2013} for life expectancy. Between-country correlation in
fertility rates is accounted for by the method of \citet{FosdickRaftery2014},
and correlation between male and female life expectancy is modeled
by the method of \citet{RafteryLalic2014}.

These statistical models allow a large
number of trajectories of future fertility and mortality for all countries
to be simulated from their predictive probability distributions. 
Each simulated trajectory of future life expectancy is converted to
age-specific mortality rates using a modified Lee-Carter method
\citep{raftery2012, li2013, li2011, sevcikova2016}.  Future age-specific fertility and mortality rates are then converted to future population by age and sex by the 
standard cohort-component method \citep{whelpton1936,preston2001},
using the {\tt bayesPop} R package \citep{sevcikova2013}.
The result is a sample of possible future trajectories of world population
by country, age and sex, to 2100. It can be viewed as a large number of
possible versions of the  2100 revision of the 
UN's {\it World Population Prospects} (which won't exist for another
84 years).

The UN's new probabilistic {\it World Population Prospects} currently treats
countries with generalized HIV/AIDS epidemics (defined as having
had HIV prevalence greater than 2\% at any time since 1980,
and hereafter referred to as the ``AIDS countries'') differently from others.
These countries have very different mortality patterns compared to countries without
generalized HIV epidemics, with a large mortality hump in middle adulthood due to AIDS,
and so they cannot be modeled using, for example, standard model 
life tables. 

The UN's current method for probabilistically projecting mortality 
for these countries starts with the deterministic projection from 
the 2015 revision of the {\it World Population Prospects} \citep{unpd2015}. 
This uses the Spectrum model developed for UNAIDS 
\citep{spectrum2014-software, stanecki2012, stover2012}, a complex
multistate model that divides adults into 15 compartments
according to their HIV status, and models transitions among these
15 compartments. In a second stage, the Bayesian hierarchical model
of \citet{raftery2013} for life expectancy and the modified Lee-Carter method are applied to 
all countries, including the AIDS countries, yielding simulated age- and 
sex-specific mortality rates. In a third and final stage, 
the predictive distribution of each future age- and sex-specific mortality
rate is adjusted so that its median coincides with the deterministic
projection from the first stage.

This method has several limitations. It relies on the 
Spectrum multistate model, which  was developed primarily to answer
short-term policy questions about the AIDS epidemic, such as future need for
antiretroviral therapy (ART) drugs, the number of AIDS orphans, and so on.
Because of its complexity and reliance on a large number of assumptions about
transition rates between the many compartments, it was not designed for
medium and long-term projections.
Indeed the UNAIDS Reference Group on Estimation, Projection and Modelling
notes that Spectrum projections more than five years into
the future are unreliable \citep[p. 9]{spectrum2015-guide}. 

The UN Population Division (UNPD) has very different needs. It provides 
projections of population that are long-term, to 2100, but output
only overall population and vital rates, and so do not need the level of detail
in Spectrum. The need to use a different methodology for a subset of
countries (about 20\% of the world's countries) is also a difficulty.

Here we propose an alternative methodology for probabilistic projection
for the AIDS countries that is simpler than the current one, 
does not require any multistate modeling, and makes use of the 
{\tt bayesPop} methodology used by the UN for other countries.
It requires probabilistic projections of overall HIV prevalence, and these
are obtained from the relatively simple non-age-structured 
Estimation and Projection Package (EPP)
\citep{Ghys&2004,Ghys&2008,ghys2010,brown2010}. Future life expectancy is projected using a 
simple time series regression model on HIV prevalence and ART coverage.
The resulting simulated life expectancies and prevalences are converted to 
age-specific mortality rates using the HIV prevalence-calibrated model life table method of 
\citet{sharrow2014}, which replicates the middle adult mortality hump 
characteristic of HIV epidemics. The population projections are then
obtained with the same {\tt bayesPop} methodology as for other countries.

The resulting method fits observed age-specific mortality data well.
In an out-of-sample validation experiment it was reasonably accurate
and provided well-calibrated probabilistic projections of aggregate
mortality and population quantities. It is simpler than the
extant Spectrum method, but still matched its projections closely
over the next 15 years. This suggests that this method may be appropriate
for probabilistic population projection for AIDS countries.

The rest of the article is organized as follows.
In Section \ref{sect-methods} we describe our new methods and the data we use.
In Section \ref{sect-results} we give results for four countries
with generalized HIV/AIDS epidemics, chosen to represent the range of 
experience, namely Botswana, Zimbabwe, Mozambique and Sierra Leone.
Then in Section \ref{sect-validation} we give the results of an
out-of-sample validation experiment to assess our method, and we also
compare the resulting projections with those of Spectrum.
We conclude in Section \ref{sect-discussion} with a discussion of the
strengths and limitations of our proposed approach.

\section{Methods and Data}
\label{sect-methods}
 Population projection involves combining
future values of age-specific fertility, age- and sex-specific mortality, and international migration rates, in this case using the standard cohort component model.
To make probabilistic projections for high-HIV prevalence counties, we follow the procedure described by \cite{raftery2012} for making probabilistic projections of fertility in countries without generalized HIV/AIDS epidemics and use the UNPD assumptions about international migration \cite[p. 30-31]{unpd2015b}, but we modify the mortality component to account for HIV/AIDS mortality. 

\cite{raftery2012}  simulate a large number of trajectories of the Total Fertility Rate (TFR) using the Bayesian hierarchical model of \cite{alkema2011}. The projected TFRs are then converted to age-specific rates using model fertility patterns. For mortality, \cite{raftery2012} simulate an equal number of trajectories of period life expectancy at birth ($e_{0}$) for females using the model of \cite{raftery2013}. Male life expectancies are conditional on the female $e_{0}$ and are derived from a model that predicts the gap in male and female $e_{0}$ \citep{RafteryLalic2014}. These $e_{0}$ projections are converted to age-specific mortality rates using a variant of the Lee-Carter method \citep{li2013, sevcikova2016}. The fertility, mortality, and migration trajectories are then converted to a future trajectory of age- and sex-specific population values using the cohort component method \citep[ch. 6]{preston2001}. 

In our application, the projection of fertility remains the same\footnote{ Fertility is projected using the \texttt{bayesTFR} package \citep{sevcikova2011} in the statistical analysis software, R.} along with the use of the UNPD assumptions about international migration and the cohort component method to combine these trajectories. A brief description of the methods for projecting $e_{0}$ and HIV prevalence as well as converting those quantities into age-specific mortality rates follows.

\subsection{\it{Projecting HIV prevalence}}
\label{sec:prevprojmodel}
 To make probabilistic projections of $e_{0}$ for countries with generalized epidemics and to convert those projections into age-specific mortality rates, we first need projections of HIV prevalence. We use a version of the UNAIDS Estimation and Projection Package (EPP) \citep{alkema2007, brown2010, ghys2010, raftery2010} for the statistical analysis software \texttt{R} to make probabilistic projections of HIV prevalence up to 2100. 

 EPP works well to project HIV prevalence into the not too distant future, approximately 5-10 years after the latest surveillance \citep{spectrum2015-guide}, but assumptions mirroring those made by UNPD in the WPP 2010 revision were imposed on two of the EPP model parameters to make projections to 2100 \citep{unpd2011}. For most countries, the model is fitted assuming that the relevant parameters have remained constant in the past. Beginning in the start year of the projection, the parameter $\phi$, which reflects the rate of recruitment of new individuals into the high-risk or susceptible group, is projected to decline by half every 20 years. The parameter $r$, which represents the force of infection, is projected to decline by half every 30 years. The reduction in $r$ reflects the assumption that changes in behavior among those subject to the risk of infection and increases in access to treatment for infected individuals will reduce the chances of HIV transmission.\par
Then, we define the median HIV prevalence trajectory to 2100 for each country as that projected by UNAIDS and generate 1,000 trajectories from EPP to define uncertainty about that. Let $z_{c,t}$ be the HIV prevalence in country $c$ at time $t$ from the single UNAIDS trajectory. Let $z_{c,t}^k$ be the prevalence in country $c$ at time $t$ for trajectory $k$ output from EPP, $k = 1, \dots, 1000$. We create 1,000 new trajectories by multiplying the UNAIDS trajectory by scalars of the form $\dfrac{z_{c,t}^k}{z_{c,t}}$. We compute five-year averages from the new yearly trajectories to be used in the projection of female life expectancy.

\subsection{\it{Projecting life expectancy at birth, $e_{0}$}}
\label{sec:e0projmodel}
 Because a generalized HIV epidemic can have a considerable depressing effect on life expectancy at birth in a short time \citep{blacker2004, bor2013, ngom2003, obermeyer2010, poit2001, reniers2014, sharrow2013b, timaeus2004}, a model that reflects the impact of HIV prevalence and antiretroviral therapy (ART) coverage is necessary to make appropriate projections of $e_{0}$ in generalized epidemics. Let $HIV_{c,t}$ and $ART_{c,t}$ be the HIV prevalence and adult ART coverage in country $c$ at time $t$, respectively. Then, $HIVnonART(HnA)_{c,t} = HIV_{c,t} \times (100 - ART_{c,t})$, so that $HnA_{c,t}$ represents the percentage of infected people not receiving ART in country $c$ at time $t$.  The model for projecting $e_{0}$, Eq. \ref{eq:e0projmodel}, predicts the five-year increase
in female life expectancy at birth from time $t-5$ to $t$ as
\begin{equation}
{\color{black} \Delta e_{0,c,(t-5:t)} = g(e_{0,c,t-5} \vert \theta^{(c)}) + \beta_{HnA} \Delta HnA_{c,t-5} + \varepsilon_{c,t-5}}
\label{eq:e0projmodel}
\end{equation}
where $\Delta e_{0,c,(t-5:t)}$ is the change in life expectancy for country $c$ at time $t-5$ to $t$, $g(e_{0,c,t-5} \vert \theta^{(c)})$ is the double logistic fitted change in life expectancy at time $t-5$ to $t$ given life expectancy at time $t-5$, and $\varepsilon_{c,t-5}$ is the error term. We construct a Bayesian hierarchical model by placing priors on all parameters in the model. (See \citet{raftery2013} for details.) 

 The double logistic term is the same as the model used for countries not substantially impacted by HIV/AIDS \citep{raftery2013} and reflects the transition from high to low mortality, which can be broken down into two processes, each represented by a single logistic function. The first process describes initial slow growth in $e_{0}$ with small improvements in mortality at low levels of $e_{0}$ resulting from gains in hygiene and nutrition followed by a quicker pace of improvement, and the second represents continuing gains from combating non-communicable diseases \citep{unpd2015b, omran2005}.  

 The Bayesian hierarchical model (Eq. \ref{eq:e0projmodel}) is estimated via Markov chain Monte Carlo simulation. For our observed data we use five-year female life expectancy estimates from 1950-2015 for all 201 countries from the WPP 2015 \citep{unpd2015}. 
We use the single UNAIDS trajectory of HIV prevalence to construct five-year averages for 1950-2015.
 We use a single trajectory of ART coverage (percentage of seropositive individuals receiving ART) for the periods 1950-2100 for each country obtained from UNPD internal tabulations \citep{unpd2011}. Note that countries not experiencing a generalized epidemic have $HnA$ values of 0 for all time periods. To project female life expectancy to 2100, we input each trajectory of $HnA$ and life expectancy period by period into the model starting from 2015, drawing random perturbations, $\varepsilon_{c,t}$, from a zero-mean Gaussian distribution whose variance is a function of life expectancy at time $t$ for each five-year time period. Countries with a generalized HIV/AIDS epidemic have more intrinsic variability than those without. Therefore, HIV/AIDS countries draw random perturbations from a distribution with larger variance than those not experiencing an epidemic. The process for deriving sex- and age-specific mortality rates from the projected female $e_{0}$ is described in the following section. 

\singlespacing \subsection{\it{Converting $e_{0}$ and HIV prevalence projections to age-specific mortality rates}}
\paragraph{} Once we have obtained probabilistic projections of HIV prevalence and  female life expectancy, we need to map those quantities onto a sets of sex- and age-specific mortality rates that can be combined with age-specific fertility rates and net migration using the cohort component method. In the WPP 2015 Revision for countries without high HIV prevalence, age-specific mortality rates are either extrapolated from recent data if they are available or obtained by converting $e_{0}$ projections to age-specific mortality rates using model mortality patterns \citep[p. 27-28]{unpd2015b}, but recent data are typically unavailable in this region of the world and model mortality  patterns are unable to replicate the particular age pattern of mortality resulting from large scale HIV epidemics \citep[p. 28]{unpd2015b}.  Available model mortality patterns also have no relationship to HIV prevalence. 

 To convert the female life expectancy and HIV prevalence projections to sex- and age-specific mortality rates we use the model of \cite{sharrow2014}, shown in  Eq. \ref{eq:compModel}. This model can reproduce the characteristic age pattern of mortality associated with generalized epidemics, i.e. an accentuated adult mortality hump concentrated at ages 30 to 45. The model represents the age pattern of mortality rates as a weighted sum of three age-varying components. The components, $b_{i,x}$ in Eq. \ref{eq:compModel}, are derived from a Singular Value Decomposition (SVD) of the matrix of observed historical mortality rates and the weights, $\omega_{i,\ell}$, are modeled as a function of HIV prevalence and female life expectancy at birth. As described in \cite{sharrow2014}, HIV MLT is not currently calibrated with ART coverage because there is not sufficient variation in country-level ART coverage for it to be a significant predictor of the weights in Eq. \ref{eq:compModel}. However, we have included the effect of ART in the life expectancy projection, and as ART coverage nears universal levels in the future, the observed calibration data will likely include sufficient variation in ART coverage to be included in future iterations of this method. Importantly, the SVD is performed on a matrix that keeps the male and female mortality rates for a given country and period together such that the $b_{i,x}$ are vectors of length 44 (two times the number of age groups). This step allows us to generate male and female mortality rates simultaneously as a function of female $e_{0}$ and HIV prevalence. Because the age-varying components, $b_{i,x}$ are fixed, the component weights, $\omega_{i,\ell}$, which are modeled as a function of HIV prevalence and $e_{0}$, are the effective parameters in this model. We refer to this model as ``HIV MLT,'' for ``HIV-calibrated model life table.''
The model is defined as follows:
\begin{equation}
\ln (m_{x,\ell}) = c_{\ell} + \sum_{i=1}^3 \omega_{i,\ell} b_{i,x} +
  \varepsilon_{x,\ell},
\label{eq:compModel}
\end{equation}
where $m_{x,\ell}$ is the period age-specific mortality rate 
for age $x$ in life table $\ell$, $c_{\ell}$ is a constant specific to life table $\ell$,
$b_{i,x}$ is the value of the $i$th component for age $x$,
$\omega_{i,\ell}$ is the weight of the $i$th component for life table $\ell$,
and $\varepsilon_{x,\ell} \stackrel{\rm iid}{\sim} N(0,\sigma^2)$ is the
error term.

 Figure \ref{fig:HIVMLTfit} plots the fit from the HIV MLT model and four existing model life table systems for South Africa females 2010-2015. HIV prevalence has remained high in South Africa and was roughly 17.4\% during this period, resulting in a large adult mortality hump. Figure \ref{fig:HIVMLTfit} demonstrates how this pattern is fully captured only by the HIV MLT model. All other systems tend to produce patterns of mortality rates that match the overall level of mortality as measured by period life expectancy at birth but miss the age-specific rates, which are critical for accurate population projection. 

\begin{figure}[!h]
        \centering
        \begin{subfigure}[b]{0.475\textwidth}
     \includegraphics[width=1\textwidth]{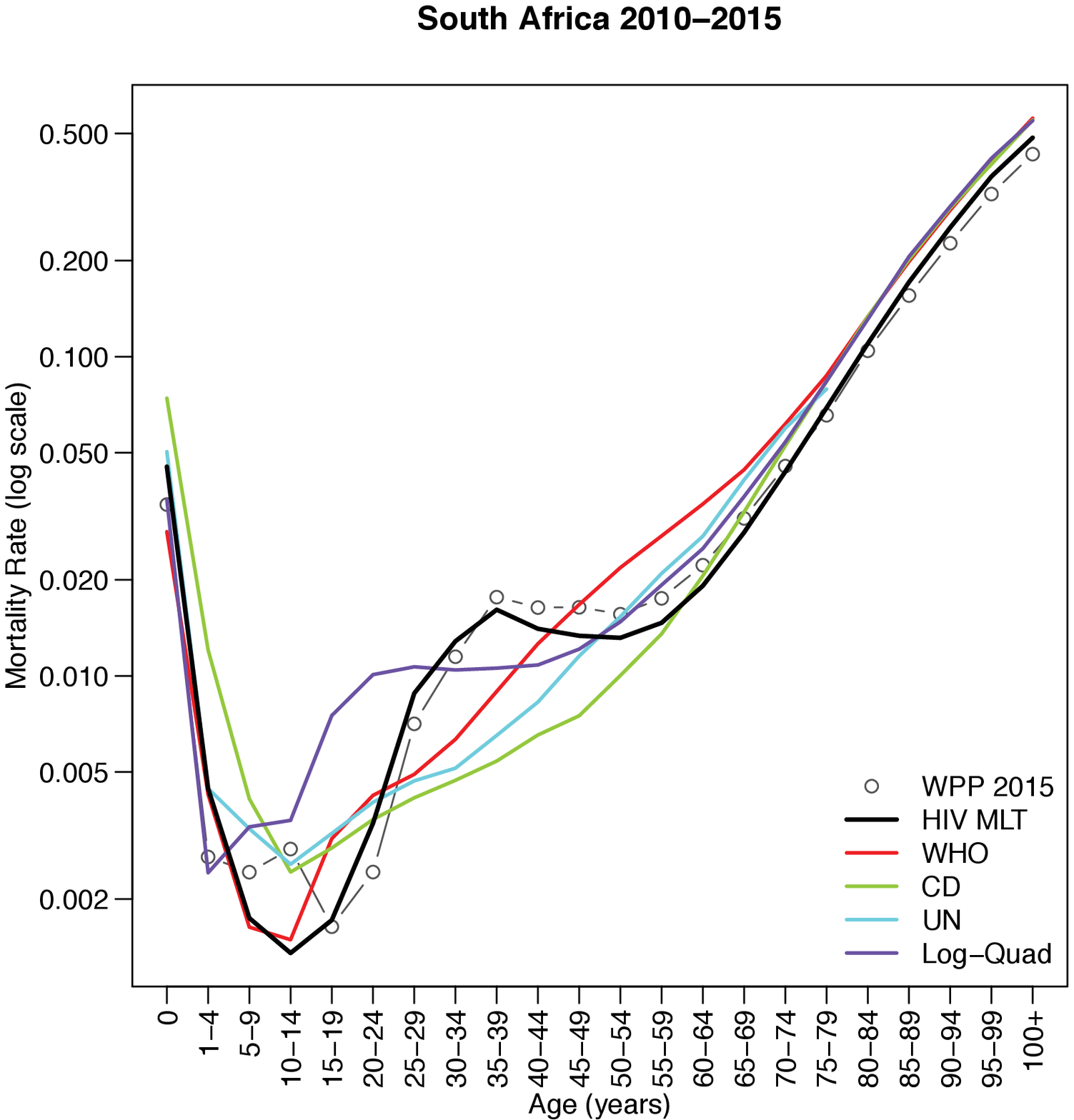}
                \caption{\small Fit of HIV MLT model  for South Africa females 2010-2015}
                \label{fig:HIVMLTfit}
        \end{subfigure}
        ~
        \begin{subfigure}[b]{0.475\textwidth}
	\includegraphics[width=1\textwidth]{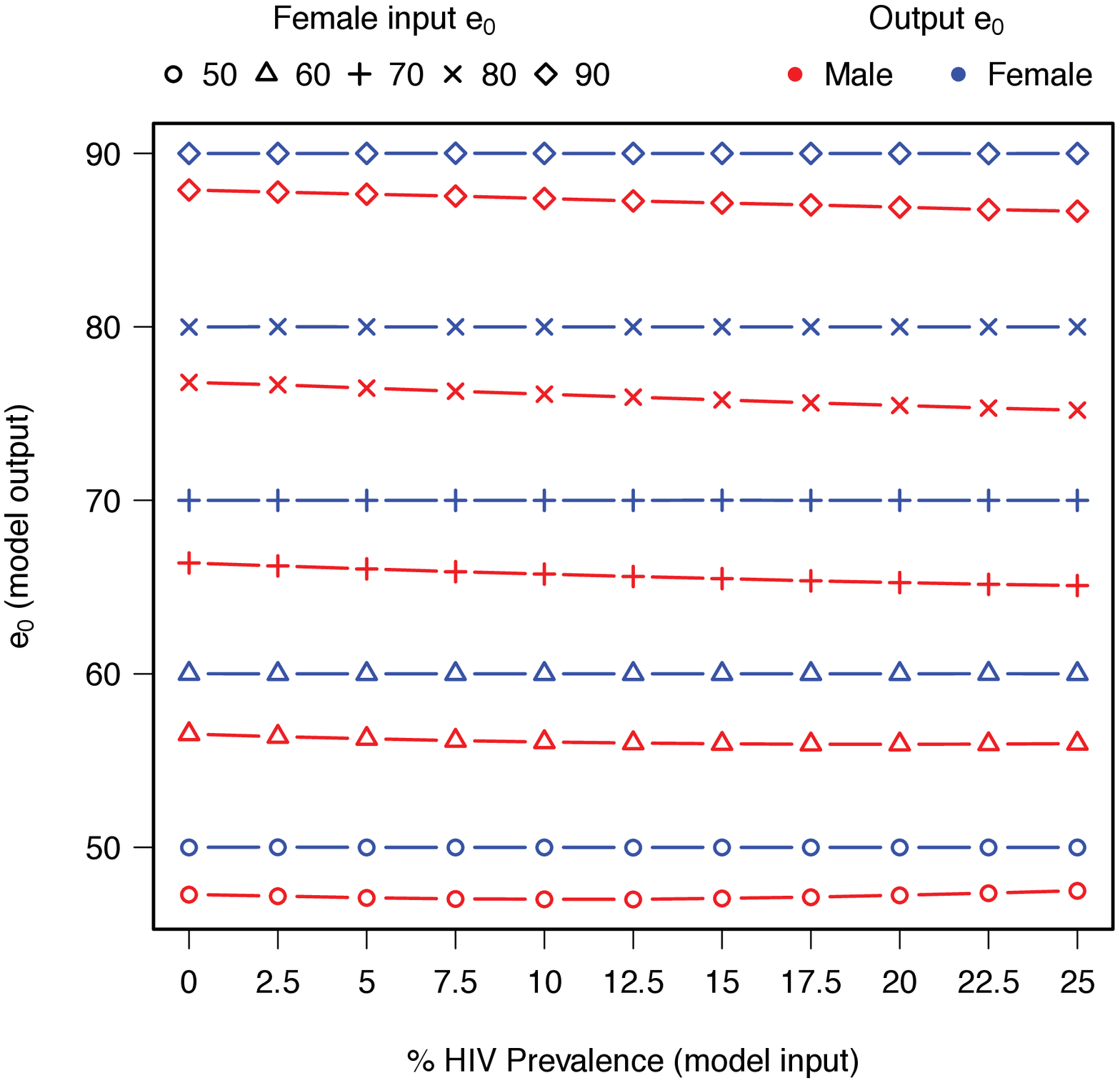}
                \caption{\small Sex-specific $e_{0}$ output for HIV MLT model at varying female $e_{0}$ and prevalence}
                \label{fig:outpute0HIVMLT}
        \end{subfigure}
        \caption{Output from HIV MLT model. (a) Fit of HIV MLT model for South Africa females 2010-2015. HIV MLT model shown with black line. For comparison, fits from the WHO modified logit model \citep{murray2003} [red solid line], Coale and Demeny model life tables \citep{coale1966, coale1983} [green solid line], UN model life tables for developing countries \citep{un1982} [teal solid line], and the Log-Quad model \citep{wilmoth2012} [purple solid line] are also shown. (b) Sex-specific output $e_{0}$, varying the two HIV MLT model inputs: female $e_{0}$ and HIV prevalence.}
        \label{fig:HIVMLTfit}
\end{figure}

 This model takes $e_{0}$ and HIV prevalence as inputs and produces a set of age-specific mortality rates that reflect those two inputs.\footnote{As described above, HIV MLT itself is deterministic (i.e. for a given $e_{0}$ and prevalence, HIV MLT will produce a specific set of mortality rates), but in validation experiments we found that prediction intervals (derived after running the probabilistic $e_{0}$ and prevalence projections through HIV MLT) with the deterministic form occasionally did not cover mortality rates at the oldest ages. To increase coverage, we add noise from the error term by randomly sampling from a normal distribution for each age with parameters mean=0 and standard deviation estimated from the distribution of residuals at each age.} HIV MLT is designed to produce a set of age-specific mortality rates that yield an output life expectancy matching the input life expectancy. The HIV MLT model was originally calibrated with sex-specific $e_{0}$ \citep{sharrow2014}, but for the present purpose we have re-calibrated the model with the female $e_{0}$ because that is what is projected by the model described in section \ref{sec:e0projmodel}. To maintain the gap between male and female $e_{0}$, the re-calibrated HIV MLT model produces complete sets of male and female age-specific mortality rates simultaneously and matches the input $e_{0}$ to the female life expectancy derived from the output female mortality rates by adjusting the intercept, $c_{\ell}$ in Eq. \ref{eq:compModel}. The male rates are adjusted using the same adjusted intercept used to match female $e_{0}$. Figure \ref{fig:outpute0HIVMLT} plots the sex-specific output $e_{0}$ from the HIV MLT model while varying the two input parameters: female $e_{0}$ and HIV prevalence. The gap in life expectancy is maintained over all combinations of the two input parameters. The female output $e_{0}$ equals the input $e_{0}$ and the male output $e_{0}$ is always below the output female $e_{0}$.

 The HIV MLT model is calibrated with five-year age-specific mortality rates obtained from WPP 2015 from 1970-2015 for the 40 countries experiencing a generalized epidemic. HIV prevalence for calibrating this model is the same as for the model for projecting $e_{0}$ (see Section \ref{sec:e0projmodel}). 
To produce a draw from the predictive distribution of the vector of 
age-specific mortality rates, a sample from each of the predictive
distributions of $e_{0}$ and prevalence projections is first produced. 
Then a sample vector of age-specific mortality rates is produced conditionally
on the simulated values of $e_0$ and HIV prevalence using 
Eq. \ref{eq:compModel}.

\subsection{\it{Making full probabilistic population projections}}
 We modified the \texttt{bayesPop} software \citep{sevcikova2013}, which combines the fertility and mortality projections using the cohort component method, to make full population projections. The software uses the method described by \cite{raftery2012} to produce mortality projections, so the package functions were altered to include the mortality methodology described above.

\section{Results}
\label{sect-results}
\begin{figure}[p]
        \centering
        \begin{subfigure}[b]{.475\textwidth}
               \includegraphics[width=1\textwidth]{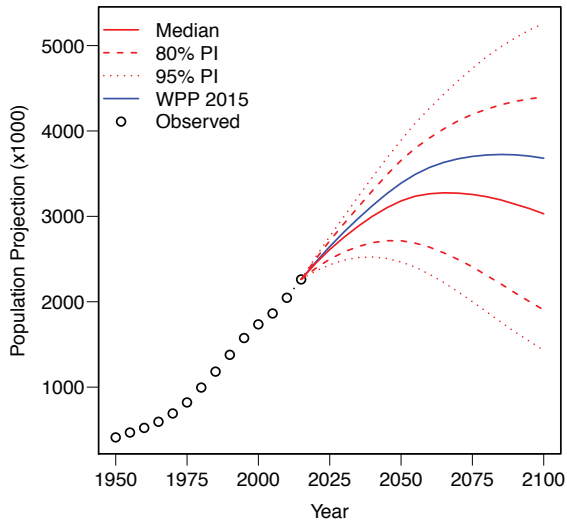}
                \caption{\small Population Projection: total population}
                \label{fig:ppBotswana}
        \end{subfigure}%
        ~ 
        \begin{subfigure}[b]{0.475\textwidth}
               \includegraphics[width=1\textwidth]{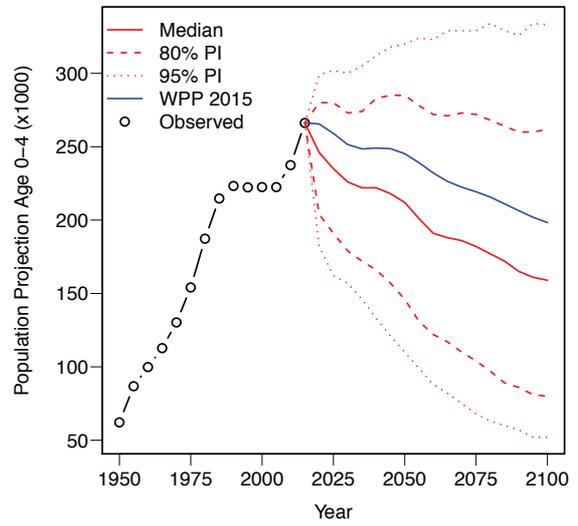}
                \caption{\small Population Projection: age 0-4}
                \label{fig:pp0-4Botswana}
        \end{subfigure}
         \\
        \begin{subfigure}[b]{0.475\textwidth}
               \includegraphics[width=1\textwidth]{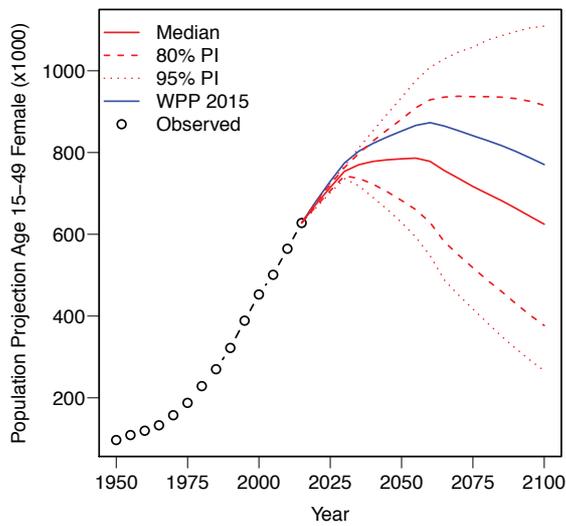}
                \caption{\small Population Projection: female age 15-49}
                \label{fig:ppF1549Botswana}
        \end{subfigure}
        ~ 
          \begin{subfigure}[b]{0.475\textwidth}
               \includegraphics[width=1\textwidth]{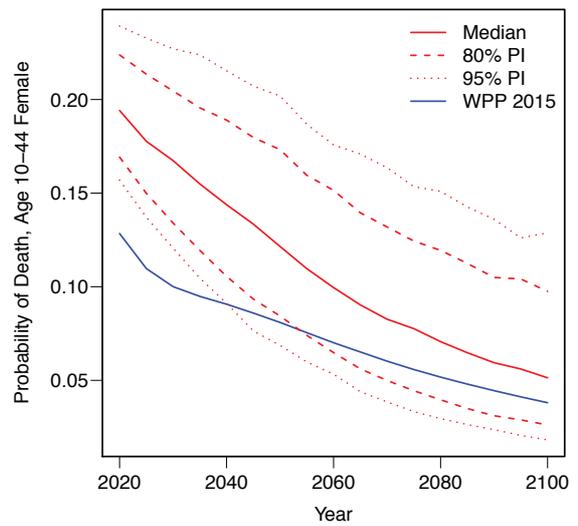}
                \caption{\small Projection of female $_{35}q_{10}$}
                \label{fig:35q10f-Botswana}
        \end{subfigure}
     
        \caption{Probabilistic population projections for Botswana 2015-2100. Past, observed data: black circles; median probabilistic projection: solid red line, 80\% predictive interval: dashed red lines; 95\% predictive interval: dotted red lines; WPP 2015 projection: solid blue line.}
        \label{fig:pppBotswana}
\end{figure}

 We discuss results here for four countries: Botswana, Zimbabwe, Mozambique and Sierra Leone. These countries were chosen to represent different levels of current HIV prevalence. Botswana and Zimbabwe represent the largest historical epidemics with peak HIV prevalences of roughly 25\%; Mozambique represents a smaller but still substantial epidemic (peak HIV prevalence $\approx$ 11\%); while Sierra Leone has a small generalized epidemic (peak HIV prevalence $\approx$ 1.6\%). Results for Botswana are shown in Figure \ref{fig:pppBotswana}, while results for the other three countries are in  Figs. S\ref{fig:pppZimbabwe}-S\ref{fig:pppSierraLeone} in Supplemental Materials.

Figure \ref{fig:ppBotswana} (top left panel) plots the total population projection for Botswana. We project an increase in total population until about 2065 when the total population begins to decline (median trajectory). Also in Figure~\ref{fig:ppBotswana} note the increasing width of the prediction intervals as the projection reaches farther into the future reflecting the increase in uncertainty, a feature of the projection for all countries. WPP 2015 (solid blue line) also shows sustained population growth followed by a mild reversal in that trend toward the end of the projection period, but our median projection predicts fewer people in the total population over the entire projection horizon compared to WPP 2015. 

\begin{figure}[!h]
        \centering
        \begin{subfigure}[b]{0.475\textwidth}
               \includegraphics[width=1\textwidth]{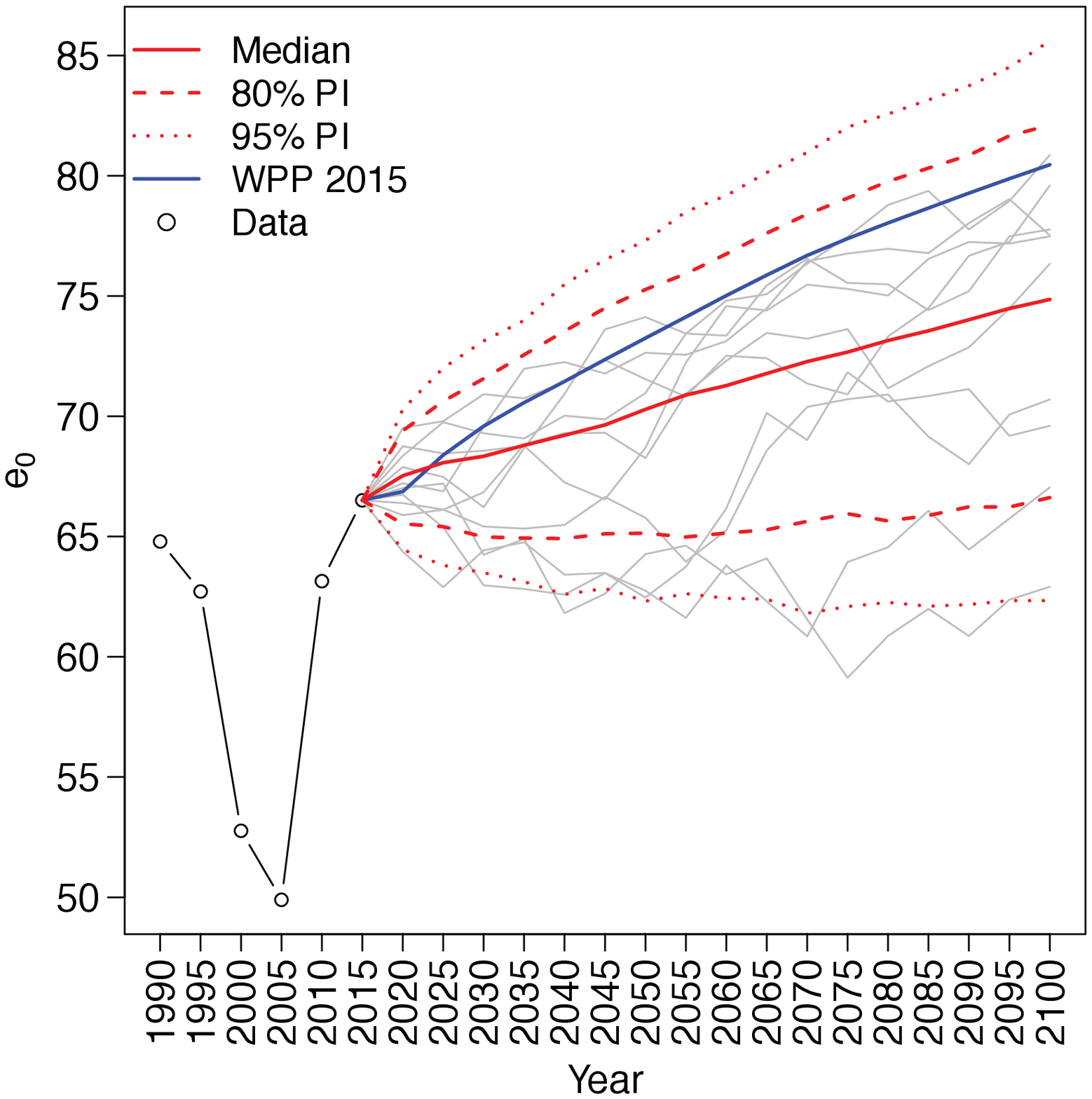}
                \caption{\small Probabilistic Life Expectancy Projection}
                \label{fig:e0projBotswana}
        \end{subfigure}
        ~
        \begin{subfigure}[b]{0.475\textwidth}
               \includegraphics[width=1\textwidth]{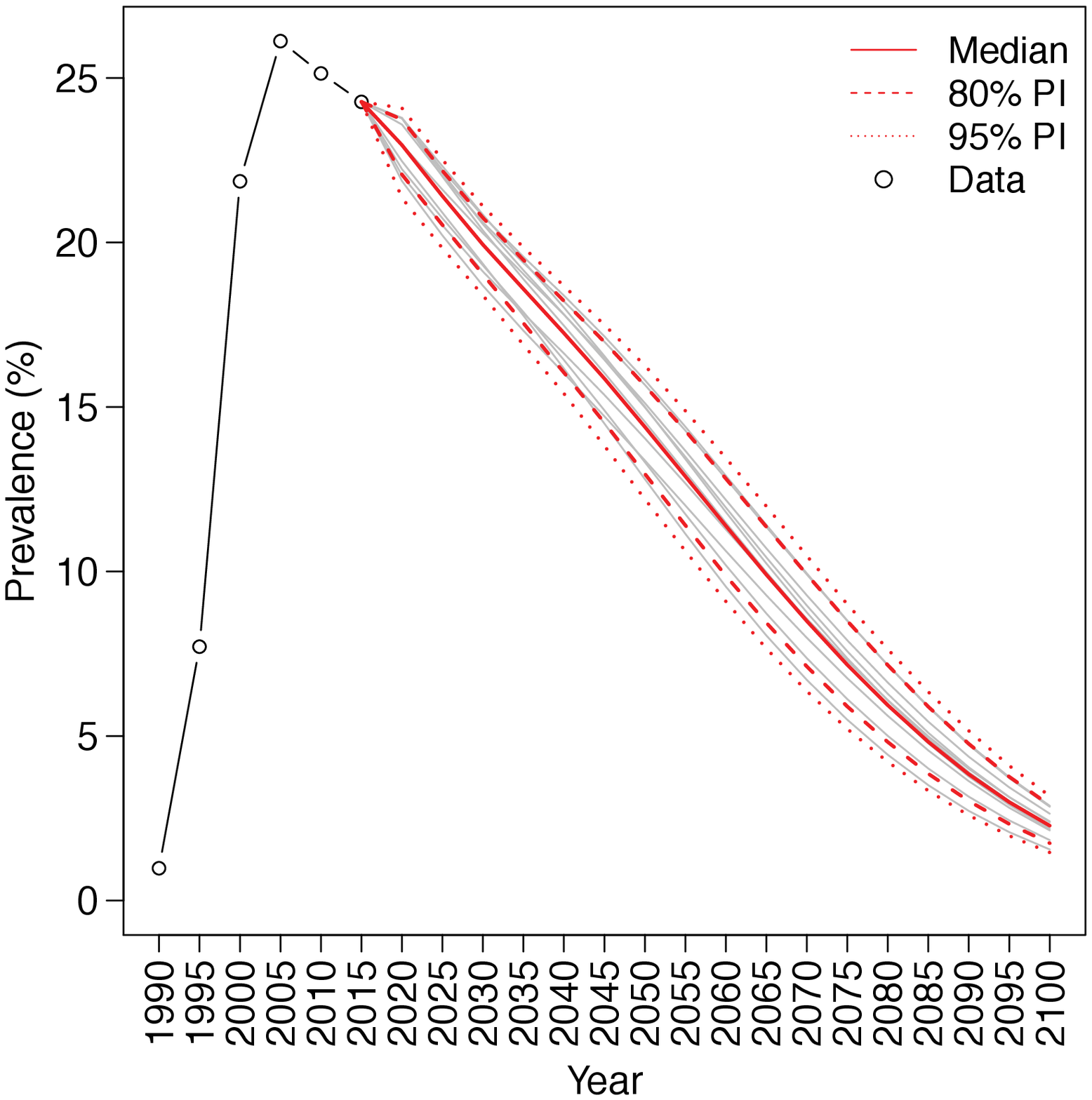}
                \caption{\small Probabilistic HIV Prevalence Projection}
                \label{fig:prevprojBotswana}
        \end{subfigure}
        \caption{Probabilistic life expectancy and HIV prevalence projections for Botswana 2015-2100. median of probabilistic projection: solid, red line; 80\% predictive interval: dashed, red line; 95\% predictive interval: dotted, red line; WPP 2015 projection: solid, blue line; observed: black circles. The gray lines in these figures are a random sample of ten trajectories from the final sample of 1,000 trajectories from the posterior distribution.}
        \label{fig:e0prevproj-Botswana}
\end{figure}

 The differences between the WPP projections and ours result mainly from our treatment of mortality, which depends on projections of HIV prevalence and life expectancy. Botswana has one of the largest HIV epidemics in the world in terms of prevalence. Figure \ref{fig:e0prevproj-Botswana} shows the probabilistic projection of HIV prevalence and life expectancy at birth for Botswana 2015-2100. HIV prevalence in Botswana is projected to remain high, dropping from a current level of about 25\% to roughly 13\% by 2050 in the median projection. Although declining, these high prevalence rates result in relatively low female life expectancies as modeled with Eq. \ref{eq:e0projmodel},
reaching just above 70 years by 2050, up from less than 50 years in 2005. Compared to the WPP 2015 projected life expectancy, our median projection shows sustained lower life expectancy with an increasing gap over the entire projection period. 

\begin{figure}[!h]
\begin{center}
\includegraphics[width=1\textwidth]{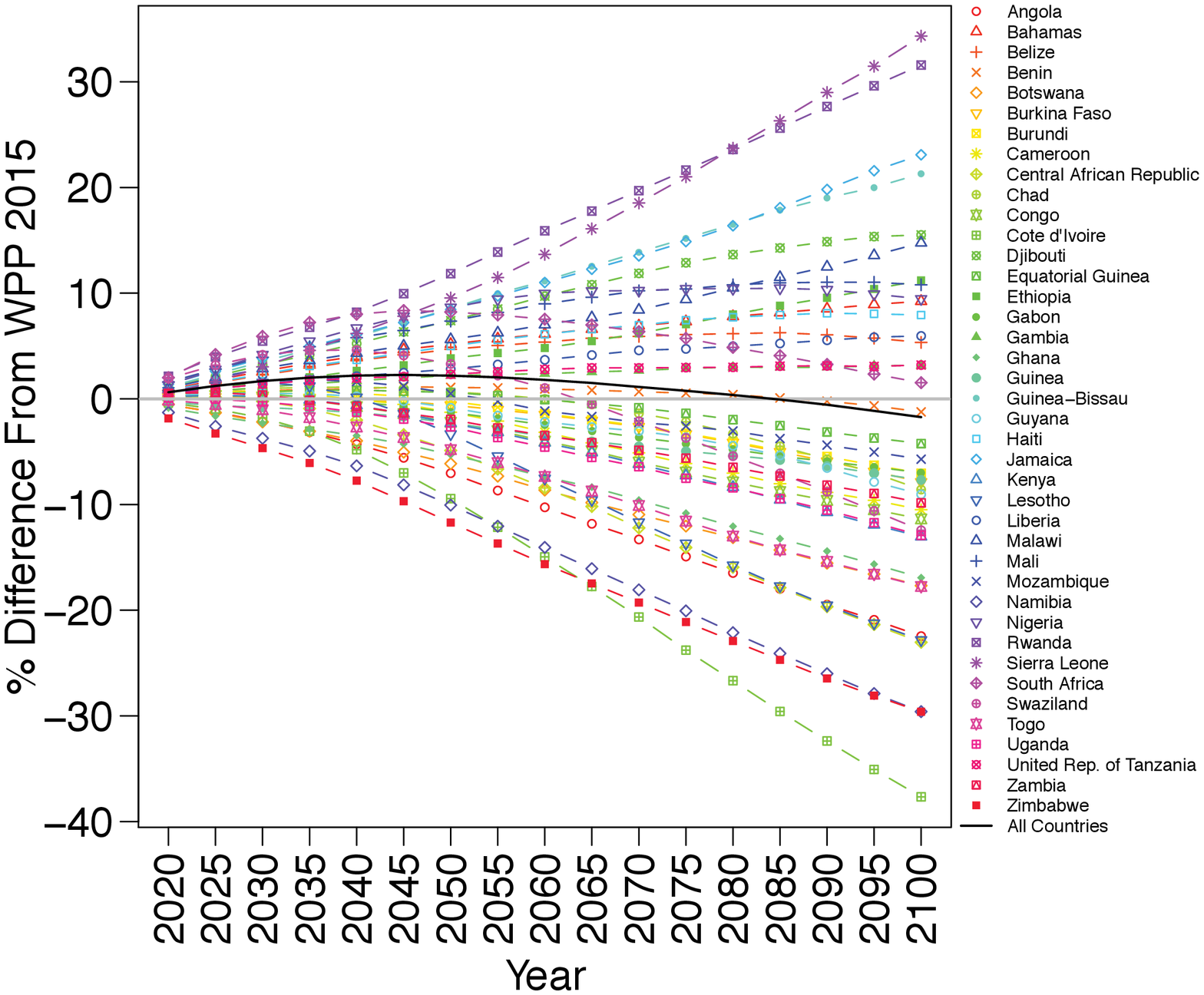}
\caption{Comparison between our median total population projection and WPP 2015 median projection for total population. Each line shows the percentage difference from the WPP 2015 total population estimate for each country and the solid black line shows the difference at each period for all countries combined.}
\label{fig:compar2050}
\end{center}
\end{figure}

Figure \ref{fig:compar2050} shows the difference between our median projection of the total population and the WPP 2015  projection of total population as a proportion of the WPP 2015 estimate from 2015-2100 for each of the 40 countries with generalized epidemics. By 2050, Botswana has one of the largest differences from WPP 2015 with about 7\% fewer people in the total population compared to the WPP 2015 Revision. In addition to projecting lower female life expectancy than WPP 2015 (see Figure \ref{fig:e0projBotswana}), our method also produces relatively high age-specific mortality rates at ages 25-45, consistent with the mortality generated under high HIV prevalence. This decline in projected total population is evidence of the crucial role age-specific mortality rates play in shaping future population makeup in countries with generalized HIV epidemics. Sustained high mortality during the reproductive years for women compared to WPP 2015 results in fewer women alive during the reproductive years and thus fewer births compared to WPP 2015.

 These effects can be seen in Figs. \ref{fig:pp0-4Botswana}-\ref{fig:35q10f-Botswana}. Figure \ref{fig:ppF1549Botswana} shows the probabilistic population projection for women aged 15-49. We project a decline in the number of women in this age category beginning around 2055, and Figure \ref{fig:pp0-4Botswana} shows a shrinking population under age five over the entire projection period. The effect of high mortality in the reproductive adult years on future population 
size, especially for women, reverberates for a number of years as smaller 
cohorts are born in each projected period, resulting in smaller total population compared to WPP 2015. Combined with declining fertility, the effect of high adult mortality yields an eventual reversal in population growth for Botswana.

Similar conditions exist for some of the other countries with large negative proportional differences at 2050 (e.g. Zimbabwe, Botswana, Namibia, Central African Republic--all with differences of greater than five percent at 2050) shown in Figure \ref{fig:compar2050}. These countries typically have large scale HIV epidemics ($> 10\%$  prevalence) and our median life expectancy projection is lower than that for WPP 2015, reducing the number of women of reproductive age resulting in smaller birth cohorts. Zimbabwe, for example, is projected to have about 13.7\% fewer people in the total population in 2050 compared to the WPP 2015 Revision. HIV prevalence is also projected to decline from about 14\% to 10\% between 2015 and 2050.  We project higher probabilities of death for women of reproductive age compared to WPP 2015 (Figure S1d) over the entire projection period resulting in a likely decreasing number of women of reproductive age past 2050 (Figure S1c) and consequently smaller birth cohorts over the projection horizon (Figure S1b).

For countries with smaller HIV epidemics, the reduction in the number of women of reproductive age is less severe and the difference between our projections and WPP 2015 tends to be smaller. Mozambique is projected to have approximately 7\% prevalence (median projection) by 2050, down from around 11\% in 2015. Figure S2a shows that our median projection of the total population is similar to the WPP 2015 projection over the entire projection period. Theses smaller differences between our projections and the WPP 2015 projections can also be seen in Figs. S2b and S2c showing the population projections for under age five and women age 15-49 respectively.  

Some countries see larger populations at 2050 in our projection compared to WPP 2015. For Sierra Leone, where HIV prevalence is projected to decrease from about 1.5\% in 2015 to less than 1\% by 2050, the difference from WPP 2015 is in the opposite direction. We project about 11.5\% more people in the total population by 2050 compared to WPP 2015. The much lower rates of HIV prevalence have a far less extreme depressing effect on total population in Sierra Leone in the long run as evidenced by Figs. S3b and S3c. In addition to relatively little effect from HIV on the age-specific mortality rates, compared to WPP 2015, we project consistently higher life expectancy over the projection period for Sierra Leone, which is consistent with the life expectancy projection for other countries for which our method projects higher populations than does WPP 2015 (e.g. Djibouti, Ethiopia, Guinea Bissau, Jamaica, Nigeria, Sierra Leone; see Figure \ref{fig:compar2050}). 

\begin{figure}[p]
        \centering
        \begin{subfigure}[b]{0.34\textwidth}
               \includegraphics[width=1\textwidth]{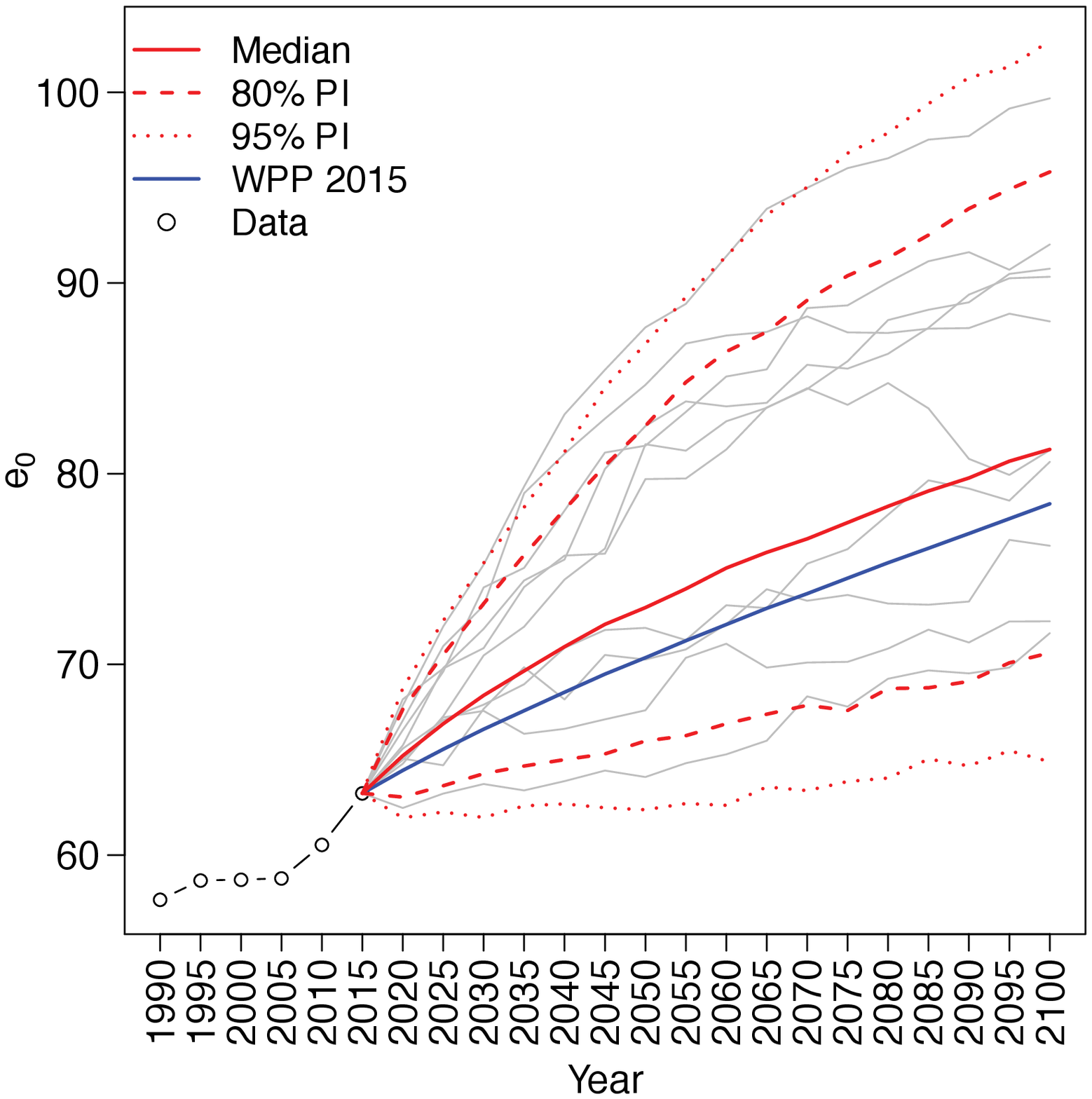}
                \caption{\small  Djibouti}
                \label{fig:e0projDjibouti}
        \end{subfigure}        
        \qquad 
        \begin{subfigure}[b]{0.34\textwidth}
               \includegraphics[width=1\textwidth]{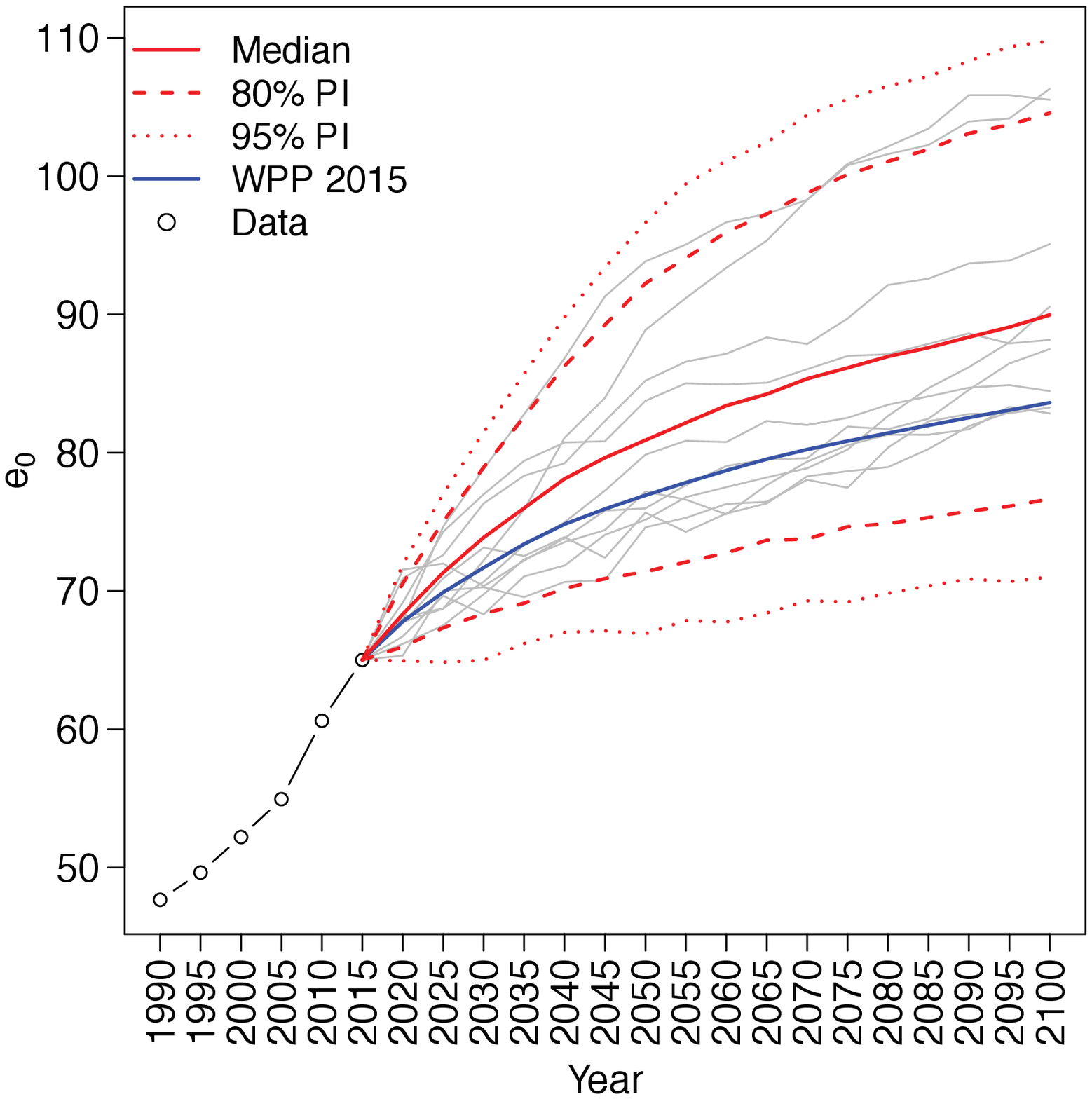}
                \caption{\small Ethiopia}
                \label{fig:e0projEthiopia}
        \end{subfigure}%
        \\
        \begin{subfigure}[b]{0.34\textwidth}
               \includegraphics[width=1\textwidth]{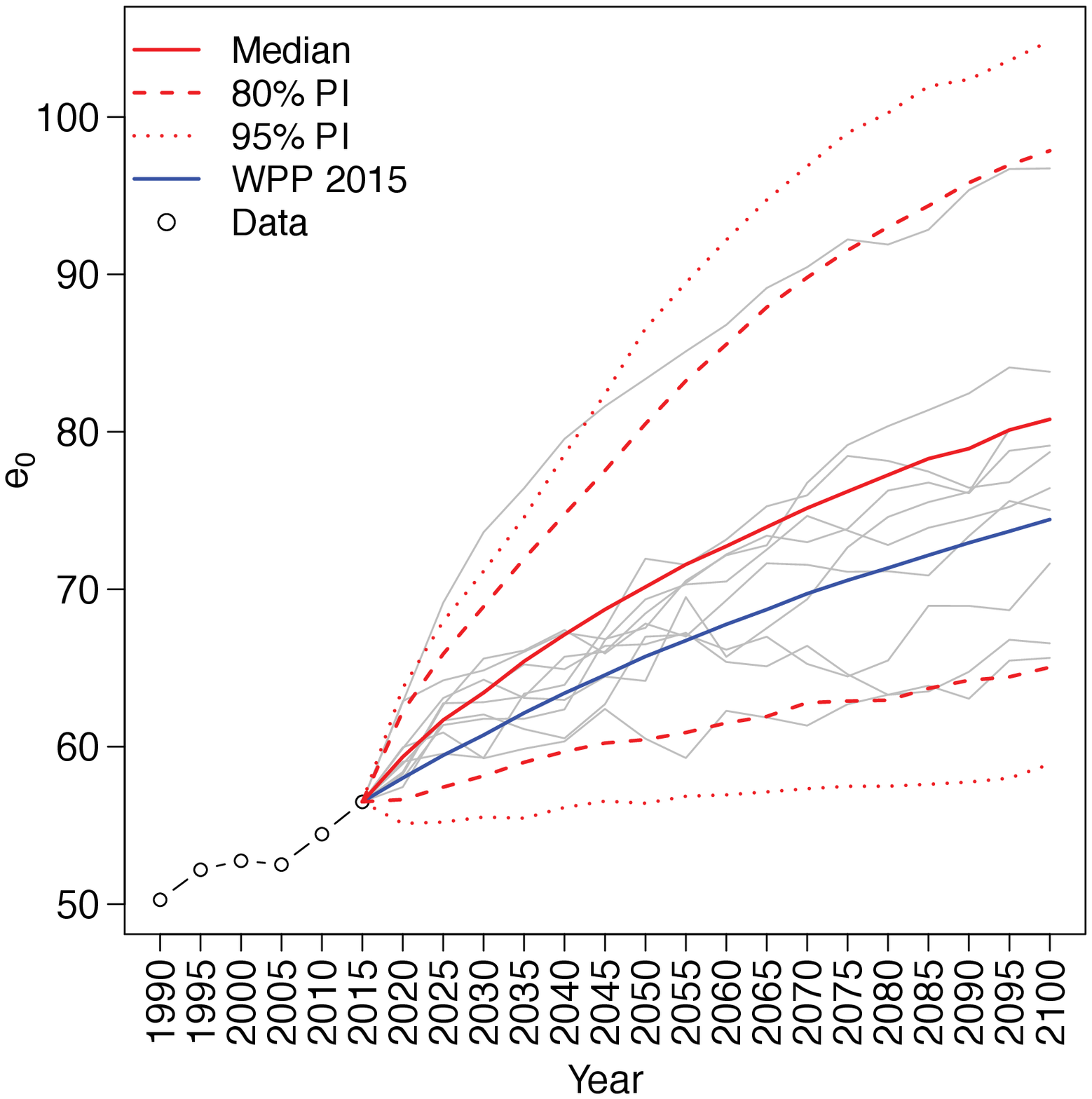}
                \caption{\small  Guinea-Bissau}
                \label{fig:e0projGuinea-Bissau}
        \end{subfigure}
         \qquad
        \begin{subfigure}[b]{0.34\textwidth}
               \includegraphics[width=1\textwidth]{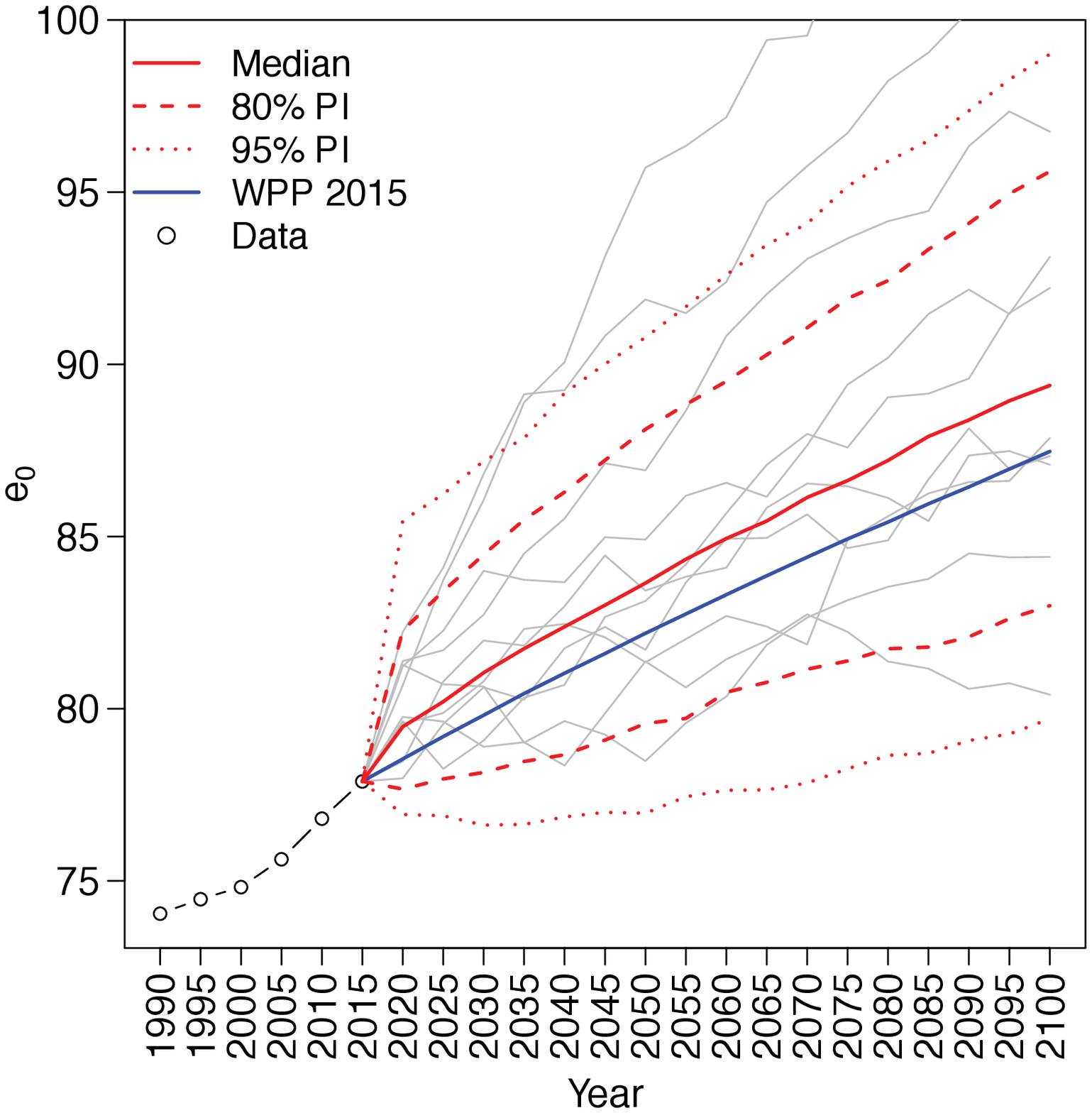}
                \caption{\small  Jamaica}
                \label{fig:e0projJamaica}
        \end{subfigure}
        \\
        \begin{subfigure}[b]{0.34\textwidth}
               \includegraphics[width=1\textwidth]{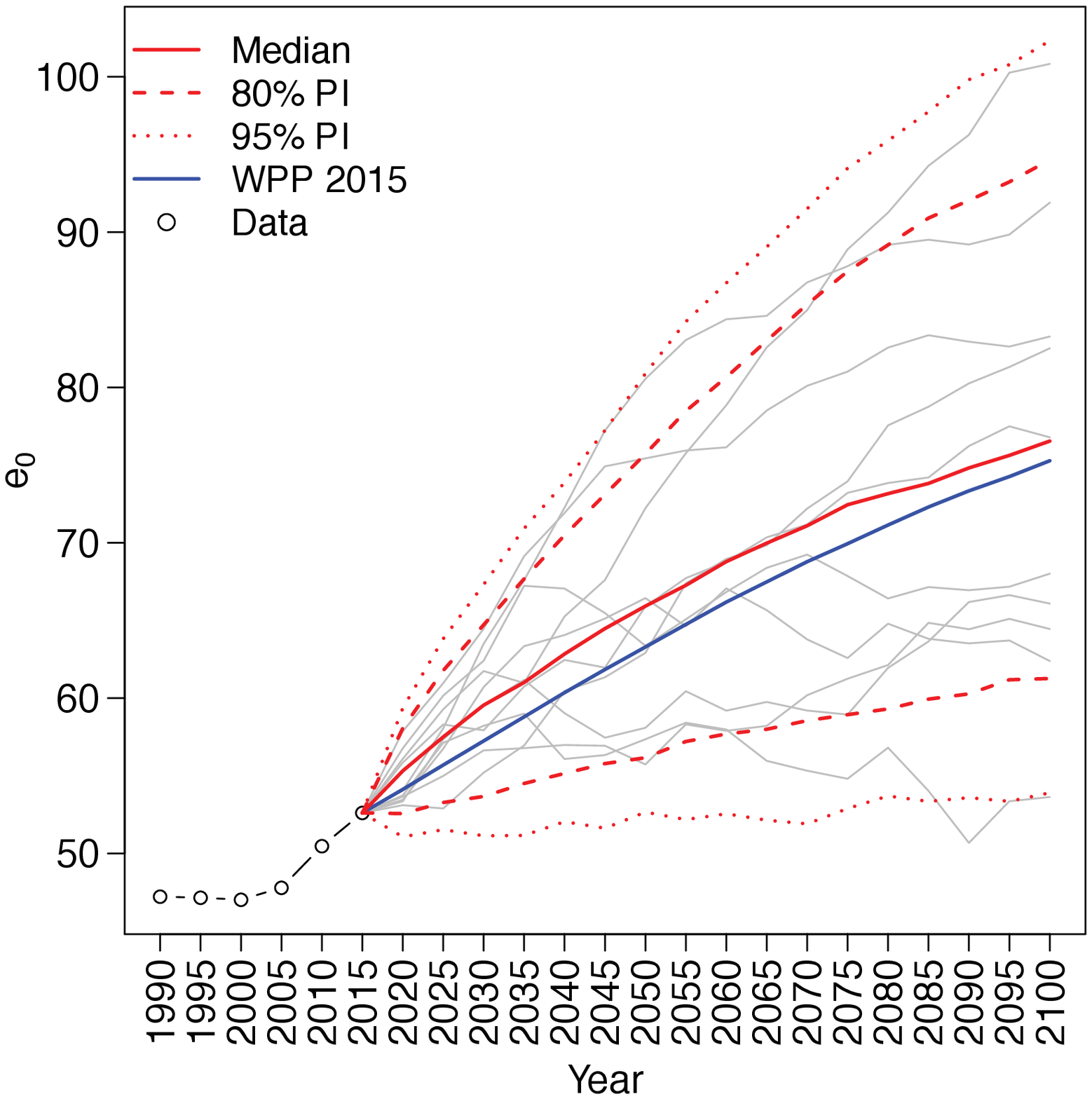}
                \caption{\small  Nigeria}
                \label{fig:e0projNigeria}
        \end{subfigure}
        \qquad 
                \begin{subfigure}[b]{0.34\textwidth}
               \includegraphics[width=1\textwidth]{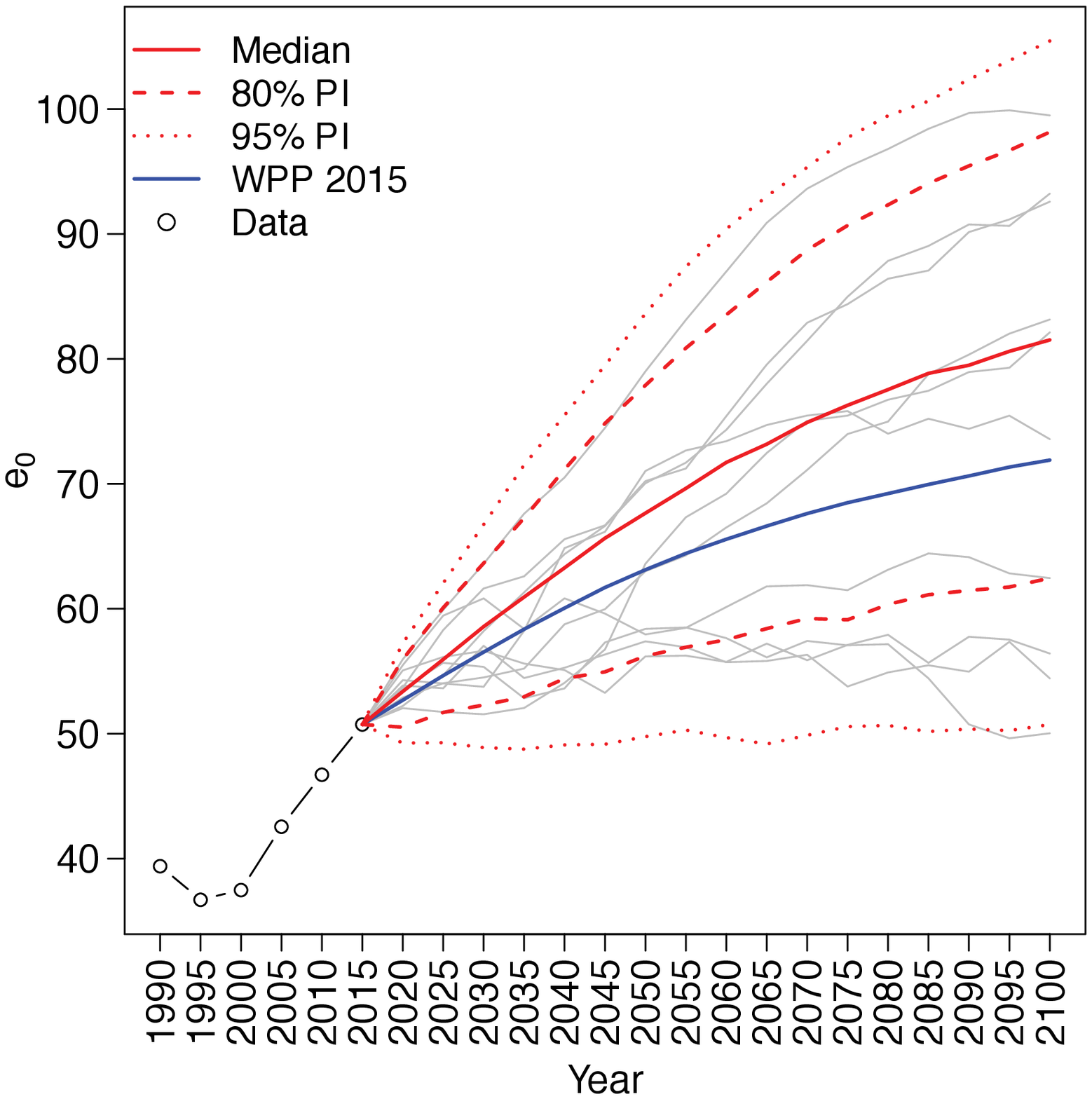}
                \caption{\small  Sierra Leone}
                \label{fig:e0projSierraLeone}
        \end{subfigure}
     
      \caption{Probabilistic life expectancy projections for six countries 2010-2100. median of probabilistic projection: solid, red line; 80\% predictive interval: dashed, red line; 95\% predictive interval: dotted, red line; WPP 2015 projection: solid, blue line; observed: black circles. The gray lines in these figures are a random sample of ten trajectories from the final sample of 1,000 trajectories from the posterior distribution.}
        \label{fig:e0projposdiff}
\end{figure}

Figure \ref{fig:e0projposdiff} plots the probabilistic projections of female life expectancy for six countries with positive differences 
in projected population at 2050 as shown in Figure \ref{fig:compar2050}. Figure \ref{fig:e0projposdiff} also depicts the WPP 2015 female life expectancy projections for these countries. As this figure shows, a large portion of the difference between the WPP 2015 total population projections and our median projections for these countries arises from differences in the projections of life expectancy, since these countries have comparatively small HIV epidemics, which will limit the influence of prevalence on the age pattern of mortality rates.  

\section{Validation}
\label{sect-validation}
\subsection{\it{Out-of-sample validation}}
 To validate our method, we calibrated the $e_{0}$ projection and HIV MLT models with five-year data from the WPP 2015 revision up to 2010 and used those models to generate a population projection for the 2010-2015 period for each of the 40 countries. We then compare the resulting mortality and population distributions with the observations from WPP 2015 for 2010-2015. 

 Because our method addresses the mortality component of the projection, we first assess the accuracy of the mortality predictions for 2010-2015 by calculating the mean absolute error (MAE) of our median projection for 2010-2015,
treating the WPP 2015 estimate as the observed value,
for four mortality indicators: $e_{0}$ (life expectancy at birth), $_{5}q_{0}$ (the probability a newborn will die before reaching age five), $_{45}q_{15}$ (the probability that a 15-year-old will die before reaching age 60), and $_{35}q_{10}$ (the probability that a 10-year-old will die before age 45). 
Table \ref{tab:mae2010-2015} presents the MAE by sex for the four mortality indicators. For males, the mean absolute error for life expectancy is about 2.3 years, while the MAE for $e_{0}$ for females is slightly higher at 2.5 years, suggesting a good fit for the level of mortality. For the other three indicators, the MAE is less than 4 per 1000 for both males and females.\footnote{Figure S\ref{fig:mort-val-dists} plots the predicted distribution of these three indicators for 2010-2015 along with the WPP 2015 estimate for the same period. This figure lends context to the magnitude of the MAEs in table \ref{tab:mae2010-2015}.} Overall, our method predicted the WPP 2015 estimate for 2010-2015 well for most countries. 

\begin{table}[h]
  \centering
  \caption{Mean absolute error for four mortality indicators for the 2010-2015 out-of-sample period ($_5q_0$, $_{45}q_{15}$ and $_{35}q_{10}$,
all x1000) using our method for projecting HIV prevalence and life expectancy and using the HIV MLT model for converting to age-specific mortality rates. All models have been calibrated with data up to 2010 and used to predict the out-of-sample period 2010-2015.}
    \begin{tabular}{rrrrr}
    \hline
    \hline
          & \multicolumn{4}{c}{Mean Absolute Error} \bigstrut\\
\cline{2-5}          & $e_0$  & $_{5}q_{0}$ & $_{45}q_{15}$ & $_{35}q_{10}$  \bigstrut\\
\hline \bigstrut[t]
    Male & \multicolumn{1}{c}{2.3}  & \multicolumn{1}{c}{16.3} & \multicolumn{1}{c}{31.9} & \multicolumn{1}{c}{22.9} \\
    Female & \multicolumn{1}{c}{2.5} & \multicolumn{1}{c}{13.9} & \multicolumn{1}{c}{33.4} & \multicolumn{1}{c}{30.6}  \bigstrut[b]\\
    \hline
    \hline
    \end{tabular}%
  \label{tab:mae2010-2015}%
\end{table}

We further assess the accuracy of our method for projecting age-specific mortality rates by calculating the proportion of age-specific mortality rates for 2010-2015 (observed data that were left out for calibration) captured in the 80\%, 90\%, and 95\% predictive intervals (PI) for 2010-2015 produced with our method. If the method is doing well, we should capture about the same percentage of the out-of-sample mortality rates as defines our predictive intervals. The top two rows of Table \ref{tab:val-coverage} show the coverage (percentage of WPP 2015 mortality rates across all ages and all 40 countries captured by the various interval widths) for the 2010-2015 period. For each PI, our method captures just under the expected proportion, indicating that the method provides reasonably well calibrated projections for the out-of-sample period. The third row of table \ref{tab:mae2010-2015} gives the coverage for the single quantity of female life expectancy in the out-of-sample period (i.e. the percentage of countries whose PIs captured the WPP 2015 estimate for female life expectancy). Coverage for female life expectancy is also good with our method again capturing approximately the expected proportion of predicted out-of-sample life expectancies. 

\begin{table}[h!]
  \centering
  \caption{Coverage of WPP 2015 sex-age-specific mortality rates, female life expectancy at birth, and total population in the 80\%, 90\%, and 95\% PIs for the 2010-2015 out-of-sample period using our method for projecting HIV prevalence and life expectancy and using the HIV MLT model for converting to age-specific mortality rates. Fertility and migration were projected using the \texttt{bayesPop} software. 
Data from 1970-2010 were used for calibration, and the resulting estimated
model was used to predict the out-of-sample period 2010-2015.}
    \begin{tabular}{rccc}
    \hline
    \hline
          & 80\% PI & 90\% PI  & 95\% PI \bigstrut\\
  \cline{2-4}   \multicolumn{1}{l}{$m_{x}$, Male} & 75    & 82    & 88 \bigstrut[t]\\
    \multicolumn{1}{l}{$m_{x}$, Female} & 69    & 81    & 86 \bigstrut[b]\\
   \multicolumn{1}{l}{$e_{0}$, Female} & 75    & 88    & 90 \bigstrut[b]\\
       \multicolumn{1}{l}{Total Population} & 97    & 97    & 100 \bigstrut[t]\\
    \hline
    \hline
    \end{tabular}
    \label{tab:val-coverage}
\end{table}

Finally, we calculate coverage similar to that of mortality and life expectancy but for the total population prediction at 2010-2015.\footnote{Calculation of total population includes running fertility projections for the out-of-sample period 2010-2015 with the \texttt{bayesTFR} software.} The bottom row of Table \ref{tab:val-coverage} shows the percentage of WPP 2015 total population estimates for 2010-2015 captured in the 80\%, 90\%, and 95\% predictive intervals. 
While the coverages for the mortality rates were not markedly different from
the nominal coverages of the PIs, coverage of the single total population estimate is near universal (i.e. the observed total population estimate for the out-of-sample period 2010-2015 was within the 80\% and 90\% PI for all but one country, Swaziland, and within the 95\% PI for all countries). 

\subsection{\it{Comparison to Spectrum/EPP}}
 Spectrum/EPP \citep{spectrum2014-software, stanecki2012, stover2012} provides useful information to UNAIDS about high-HIV prevalence countries including estimates of HIV prevalence, total population, life expectancy, and a host of other demographic and epidemiological variables in the short term (five years). However, it is, of necessity, quite complex. We propose a simpler method to project over a longer projection horizon, but it is instructive to compare our projection results to those obtained from Spectrum in the near term. 

Table \ref{tab:val-spec} shows the mean difference (MD) between our median projection using the full dataset for calibration (i.e. no out-of-sample period was removed) and Spectrum among all 40 countries for prevalence and life expectancy at birth for the first three projection periods.\footnote{All Spectrum results were obtained with Spectrum version 5.441 downloaded July 2016 \citep{spectrum2014-software}.} For prevalence, our projections differed from Spectrum by less than one percentage point on average across the 40 countries during each of the first three projection periods, which is substantially less than the prediction margin of error. The mean differences for life expectancy show that on average across the 40 countries our projection is between one and two years lower than the Spectrum projection in the very near term.
Thus, overall our projections are similar to those of Spectrum, but
using a much simpler model. 

\begin{table}[h!]
  \centering
  \caption{Mean difference (MD) for prevalence and life expectancy and mean proportional difference (MPD) for total population, population aged 0-4, and female population aged 15-49, between our method and the Spectrum/EPP software for the first three projection periods. All results for Spectrum derived from the Spectrum software version 5.441.}
    \begin{tabular}{rccrccc}
    \hline
    \hline
          & \multicolumn{2}{c}{MD$^{a}$} &       & \multicolumn{3}{c}{MPD$^{b}$ (\%)} \bigstrut\\
\cline{2-3}\cline{5-7}          & Prevalence & $e_{0}$ &       & Total Pop. & Total Pop 0-4 & Female Pop. 15-49 \bigstrut\\
\cline{2-3}\cline{5-7}    2015-2020 & 0.7  & -1.76  & \multicolumn{1}{c}{} & -0.9 & -0.8  & -0.2 \bigstrut[t]\\
    2020-2025 & 0.8  &  -1.10  & \multicolumn{1}{c}{} & -1.3 & -1.0  & -0.3 \\
    2025-2030 & 0.9  &  -0.97  & \multicolumn{1}{c}{} & -1.6 & -1.2  & -0.3 \bigstrut[b]\\
    \hline
    \hline
    \multicolumn{7}{p{.885\textwidth}}{$^{a}$ Mean Difference between our estimate and that of the Spectrum software for all 40 countries. These numbers are on the scale of prevalence (percentage points) and life expectancy (years).} \bigstrut[t]\\
    \multicolumn{7}{p{.885\textwidth}}{$^{b}$ Mean Proportional Difference is the difference between our median estimate and the Spectrum estimate as a proportion of the Spectrum estimate. Varying absolute population sizes among the 40 countries necessitate taking the proportional error. MPD is in units of per 100.}\bigstrut[t] \\
    \end{tabular}
  \label{tab:val-spec}
\end{table}

 The right three columns of Table \ref{tab:val-spec} show the mean proportional difference (MPD)
between our median projection and Spectrum for three population quantities: the total population across all ages, the total population aged 0-4, and the female population aged 15-49. The MPD is the average difference between our median estimate and Spectrum as a proportion of the Spectrum estimate for a given population indicator (results in Table \ref{tab:val-spec} for MPD shown as percentages). Again, we present a simpler method but it should approximate Spectrum at least in the short run. For total population, our median estimates of total population were about one percent lower than the Spectrum estimate for the most recent projection period, 2015-2020. The MPD is between one and two percent for each of the next two periods. Our projections of the population aged 0-4 are also lower than the Spectrum result by roughly one percent. Finally, our short term projection results for the female reproductive age population are also close to the Spectrum result with less than one percent average proportional difference over all three periods. In sum, results from Table \ref{tab:val-spec} suggest our less complex method approximates the short-term projections of Spectrum for these demographic quantities reasonably closely.

\section{Discussion}
\label{sect-discussion}
 We have presented a method for making probabilistic population projections for countries with generalized HIV epidemics. We accomplish this by following the Bayesian probabilistic projection method described by \cite{raftery2012} for fertility and the international migration assumptions of the UNPD, but because of the singular nature of mortality in generalized HIV epidemics, we modify the mortality component of the projection to incorporate the future trajectory of the epidemic in terms of HIV prevalence and ART coverage. The probabilistic fertility and mortality projections and the UNPD's assumptions about future migration are combined using the cohort component method of projection. These projections are potentially useful to researchers and policy makers as this method provides a predictive distribution for population quantities of interest such as total population, life expectancy, and support ratios into the future. Our method takes into account uncertainty about future levels of mortality and fertility, the major drivers of population change, as well as uncertainty about the trajectory of HIV prevalence.  Our approach is less complex than the UNPD's current method for projecting mortality in high-HIV prevalence settings and better captures the age pattern of mortality rates during a generalized HIV epidemic. 

Results from the projections described here show that by 2050 and beyond, we project smaller total populations for about half of the 40 countries under study here compared to WPP 2015. Many of the countries with the largest negative differences in projected total population compared to WPP 2015 have large scale HIV epidemics. For these countries, we tend to project lower total life expectancy over the course of the projection period. Combined with projected high HIV prevalence, the lower life expectancies result in high age-specific mortality rates during the younger adult years, and thus fewer women of reproductive age and consequently smaller birth cohorts. Projected into the future, these trends lead to smaller total population projections compared to WPP 2015. Coupled with declining fertility, high mortality rates resulting from HIV/AIDS-related deaths produce a reversal in population growth by 2100 for some countries with very large epidemics. Overall, these trends amount to a -1.7\% difference in total population amongst all 40 countries by 2100 compared to WPP 2015, a difference of approximately 52.5 million people. 

Although the method presented here for mortality and elsewhere for fertility takes into account uncertainty about future levels of fertility and mortality, it does not include uncertainty about international migration in the future, which can be an important source of forecast errors. Likewise, the life expectancy projection model and the model used to convert $e_{0}$ projections to age-specific mortality rates are calibrated with results from WPP 2015, some of which are themselves modeled, so they reproduce only the variability in the quantities of interest contained in the WPP results. To the extent that the WPP 2015 data and results used to calibrate these models reflect the empirical reality, the models we present here should as well. Finally, as in \cite{raftery2012}, this method does not take into account random variation in the number of births or deaths, given the fertility and mortality rates.

\section*{Acknowledgements}
This work was supported by the Eunice Kennedy Shriver National Institute of Child Health and Human Development (NICHD) under Grant  R01 HD054511; under Grant R01 HD070936; and under Grant K01 HD057246.

\newpage
\bibliography{PPPforHIVcountries-ref}

\newpage
\section*{Supplemental Material}
\subsection*{Probabilistic Population Projections for selected countries}
\begin{supfigure}[p]
        \centering
        \begin{subfigure}[b]{.475\textwidth}
               \includegraphics[width=1\textwidth]{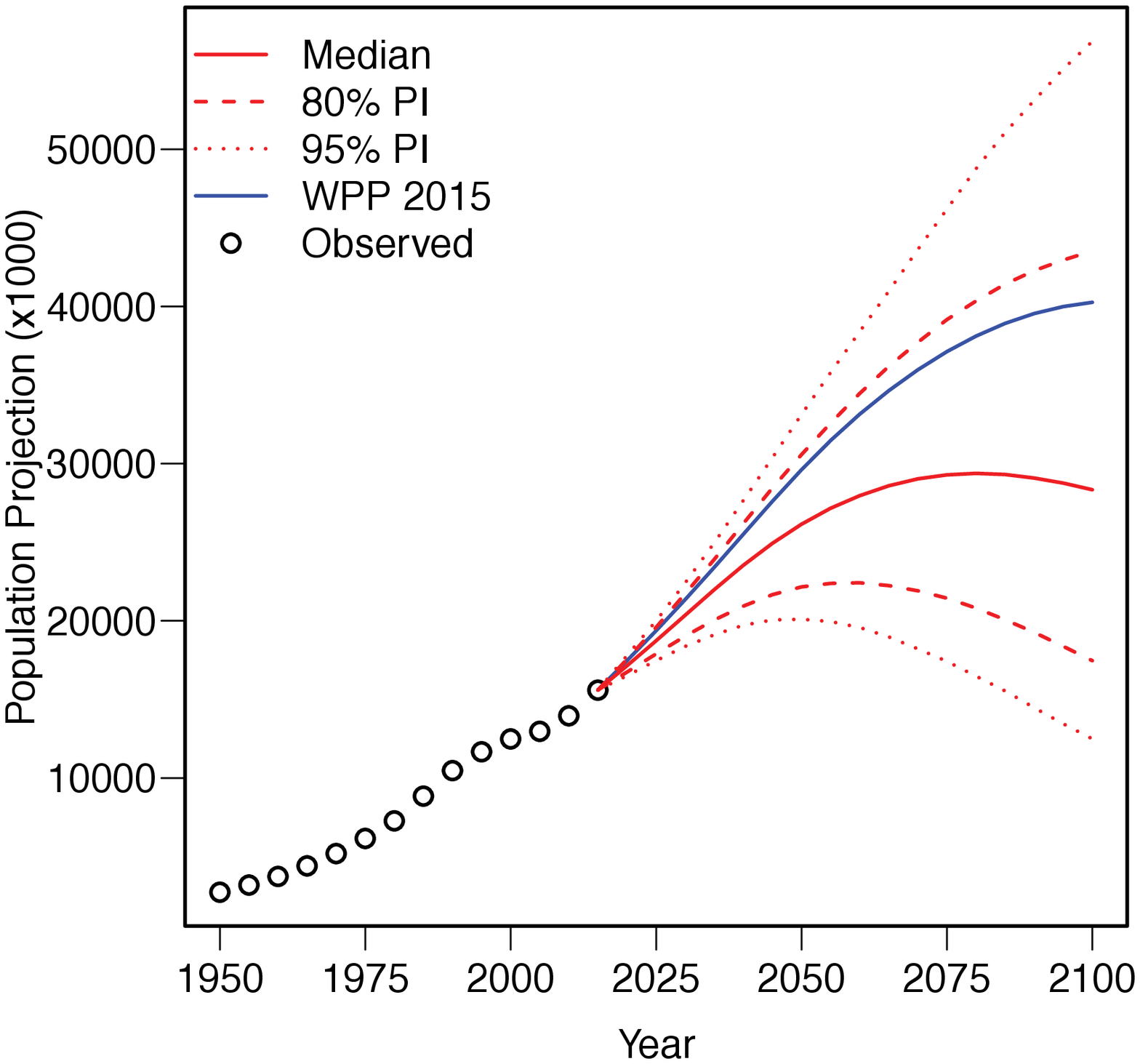}
                \caption{\small Population Projection: total population}
                \label{fig:ppZimbabwe}
        \end{subfigure}%
         ~  
        \begin{subfigure}[b]{0.475\textwidth}
               \includegraphics[width=1\textwidth]{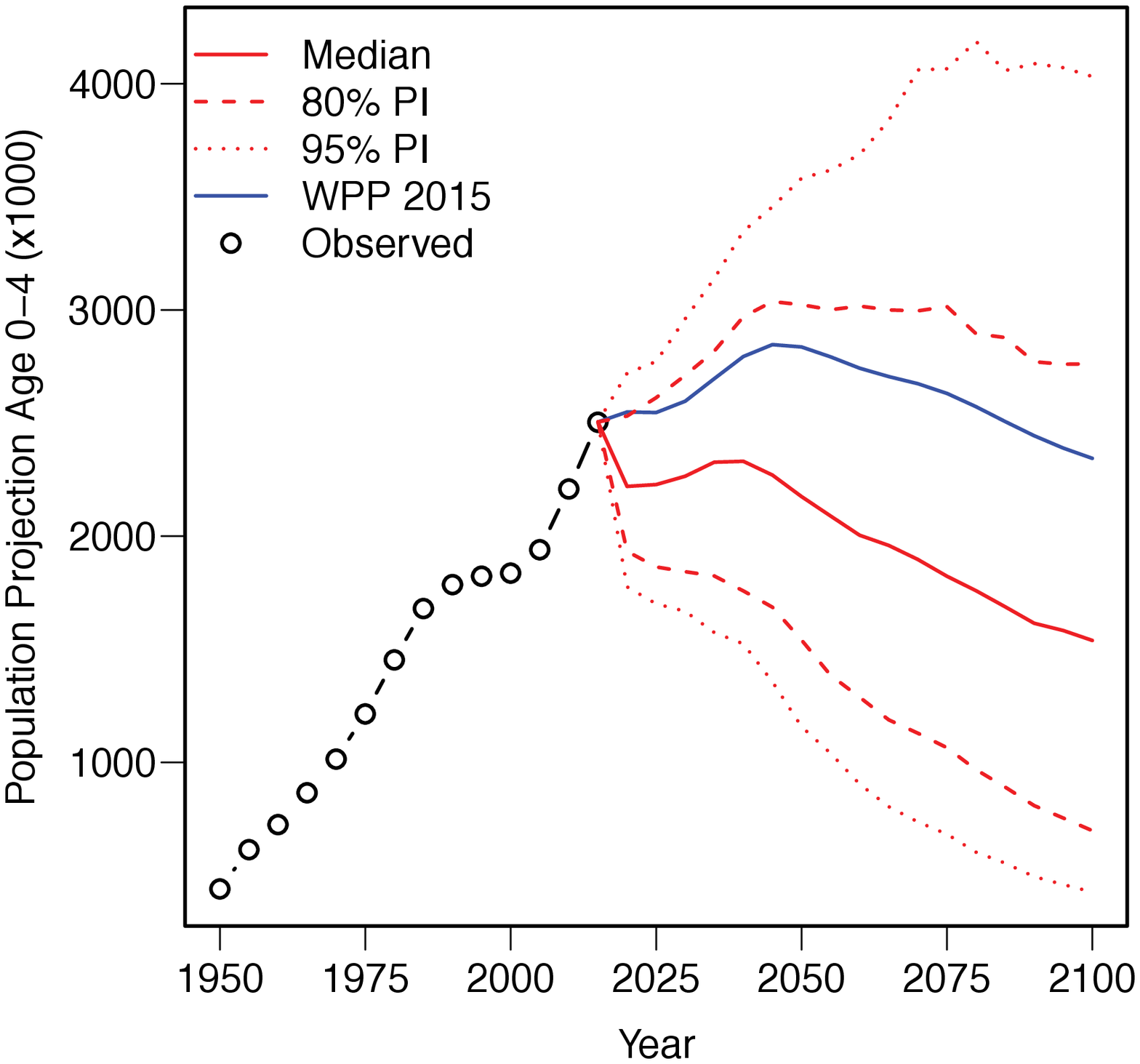}
                \caption{\small Population Projection: age 0-4}
                \label{fig:pp0-4Zimbabwe}
        \end{subfigure}
         \\
        \begin{subfigure}[b]{0.475\textwidth}
               \includegraphics[width=1\textwidth]{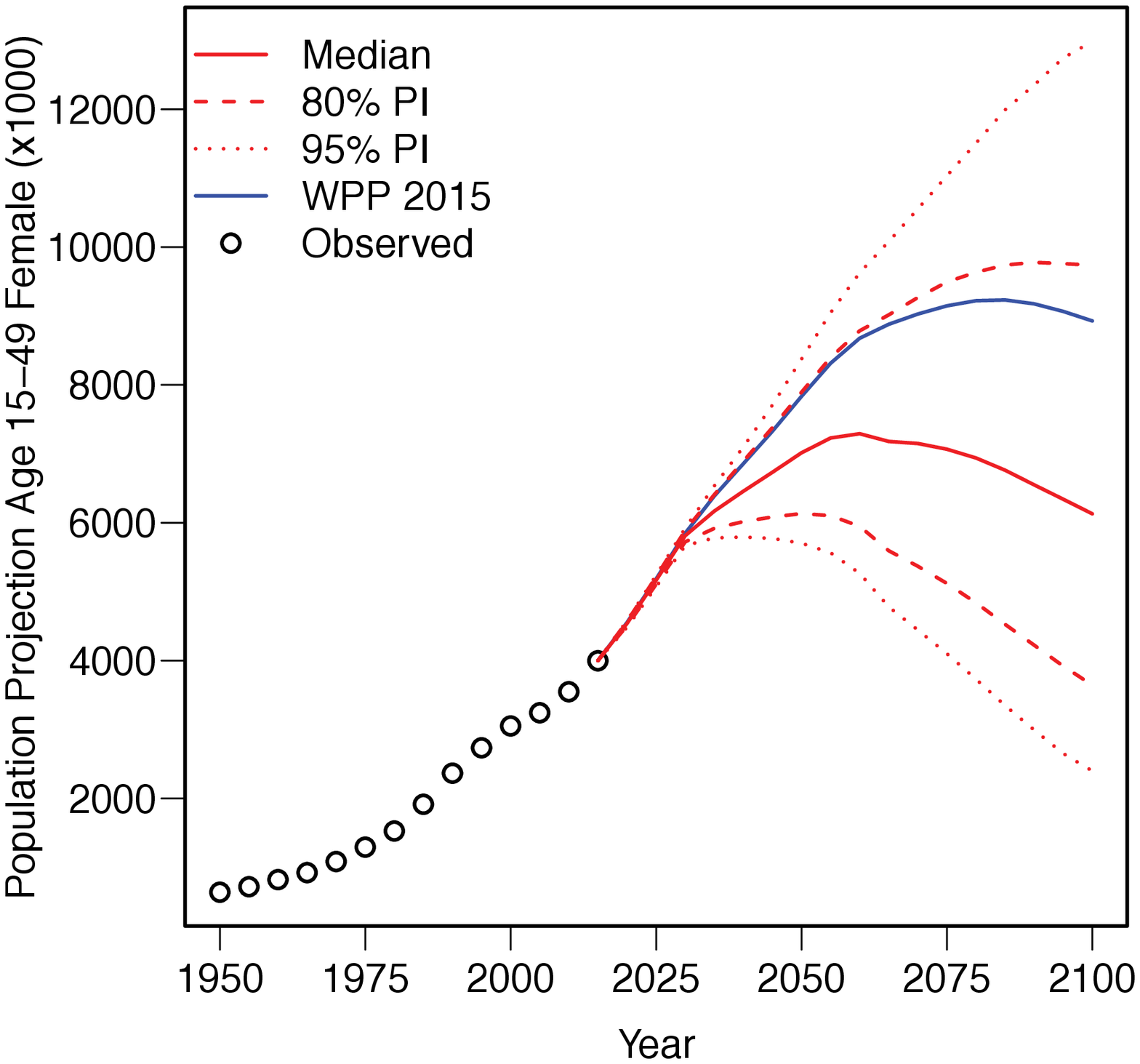}
                \caption{\small Population Projection: female age 15-49}
                \label{fig:ppF1549Zimbabwe}
        \end{subfigure}
        ~ 
          \begin{subfigure}[b]{0.475\textwidth}
               \includegraphics[width=1\textwidth]{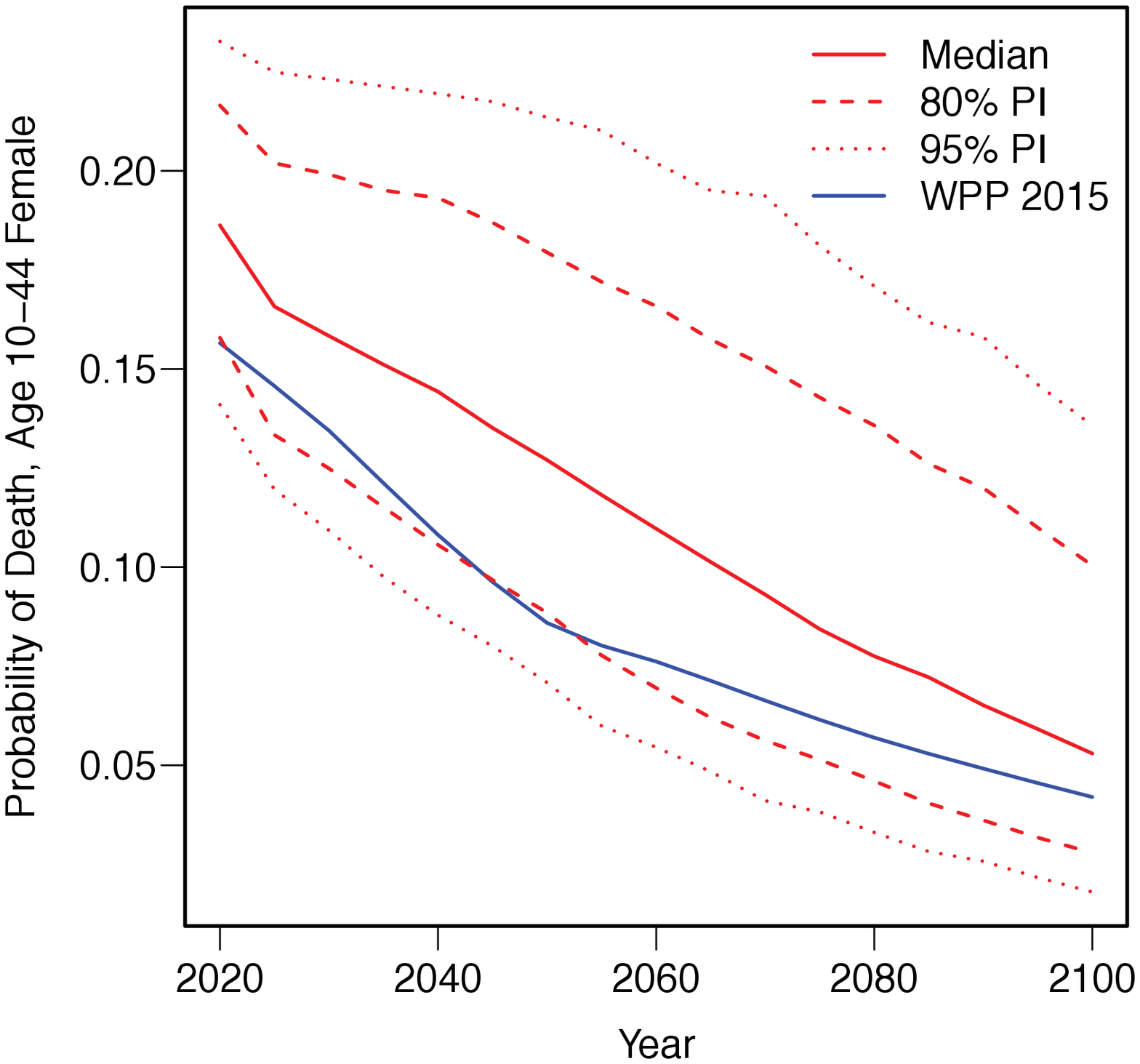}
                \caption[SM1]{\small Projection of female $_{35}q_{10}$}
                \label{fig:35q10f-Zimbabwe}
        \end{subfigure}
     
        \caption{Probabilistic population projections for Zimbabwe 2015-2100. Observed data: black circles; median probabilistic projection: solid red line, 80\% predictive interval: dashed red lines; 95\% predictive interval: dotted red lines; WPP 2015 projection: solid blue line.}
        \label{fig:pppZimbabwe}
\end{supfigure}

\begin{supfigure}[p]
        \centering
        \begin{subfigure}[b]{.475\textwidth}
               \includegraphics[width=1\textwidth]{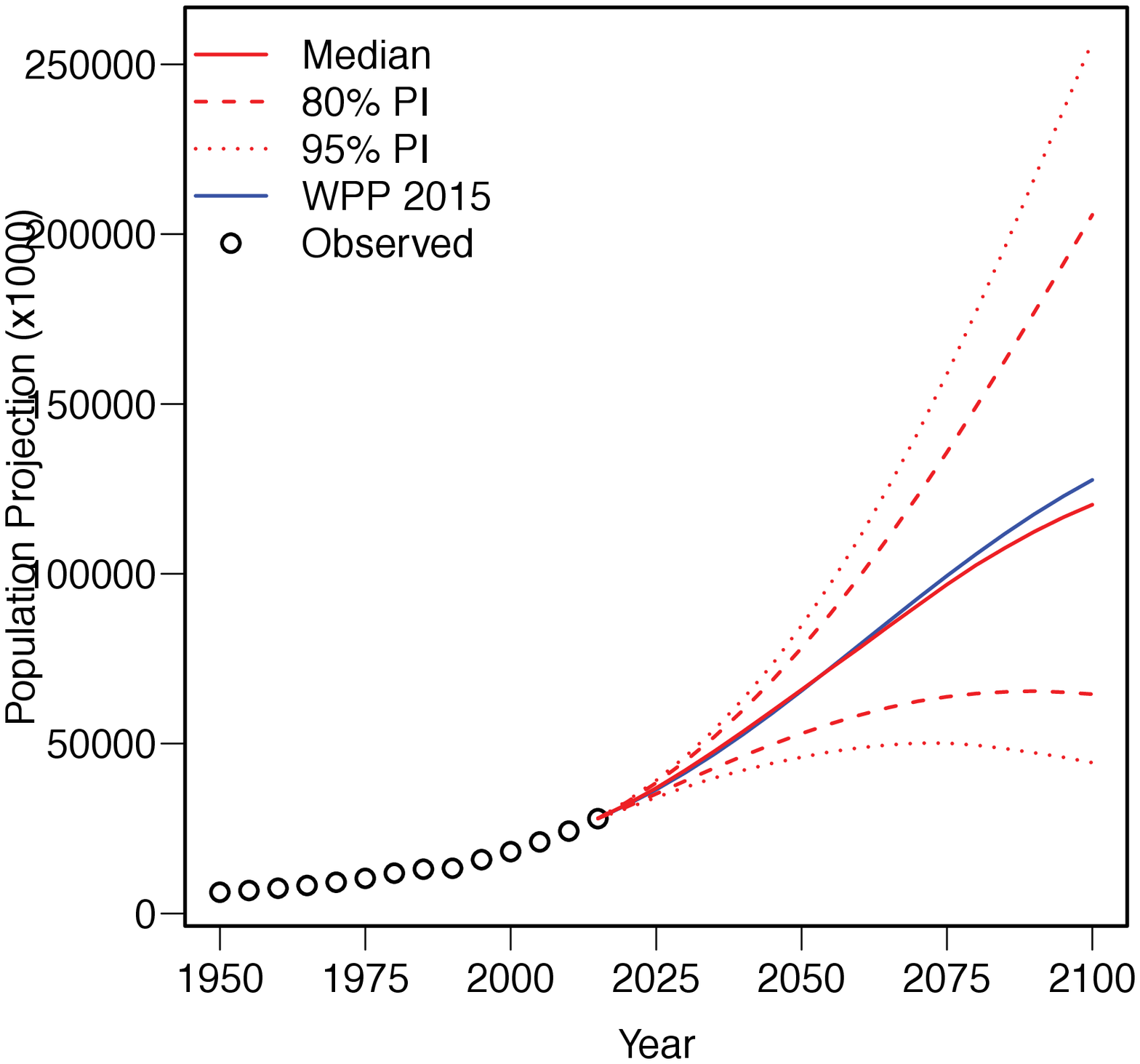}
                \caption{\small Population Projection: total population}
                \label{fig:ppMozambique}
        \end{subfigure}%
         ~  
        \begin{subfigure}[b]{0.475\textwidth}
               \includegraphics[width=1\textwidth]{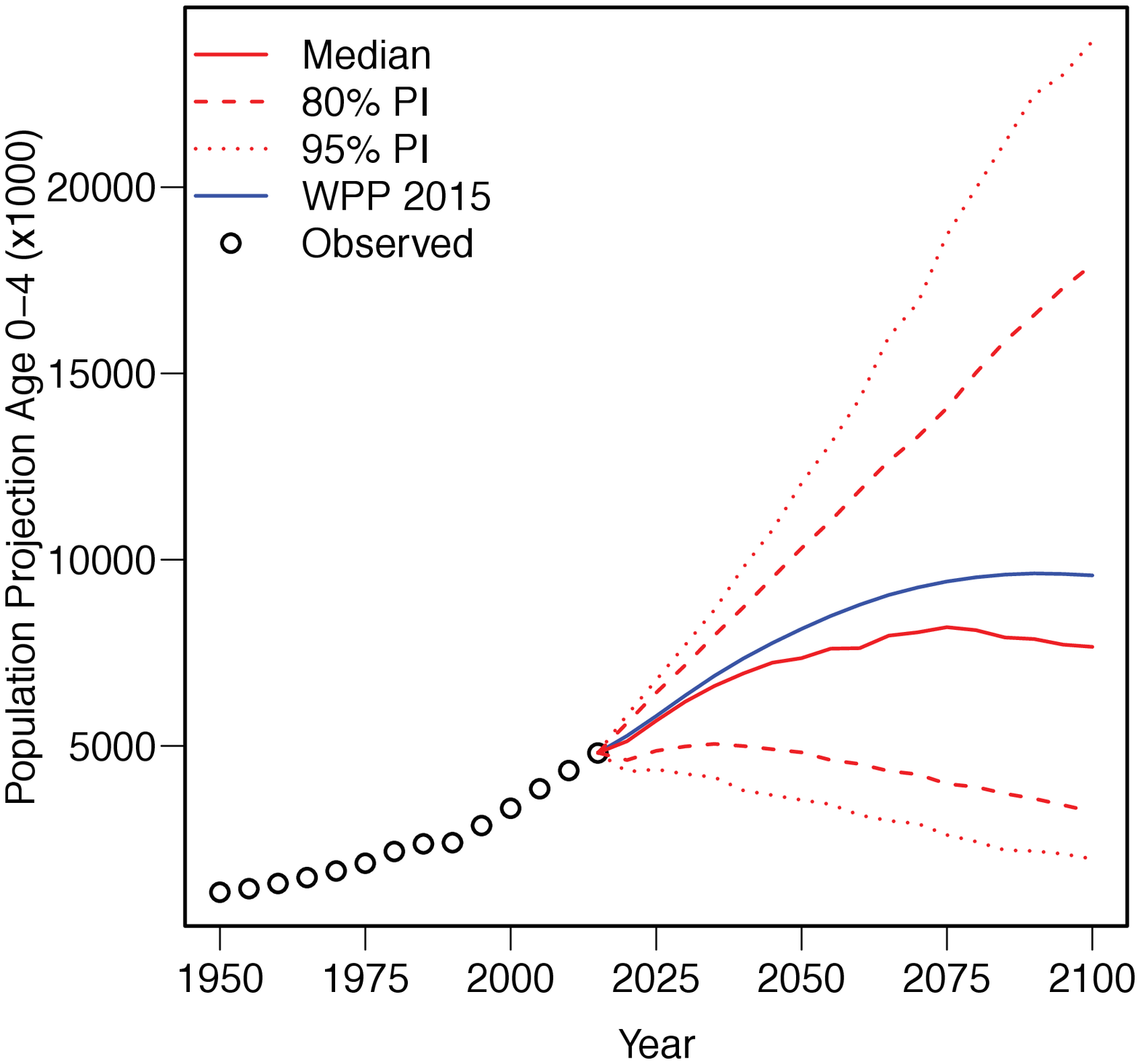}
                \caption{\small Population Projection: age 0-4}
                \label{fig:pp0-4Mozambique}
        \end{subfigure}
         \\
        \begin{subfigure}[b]{0.475\textwidth}
               \includegraphics[width=1\textwidth]{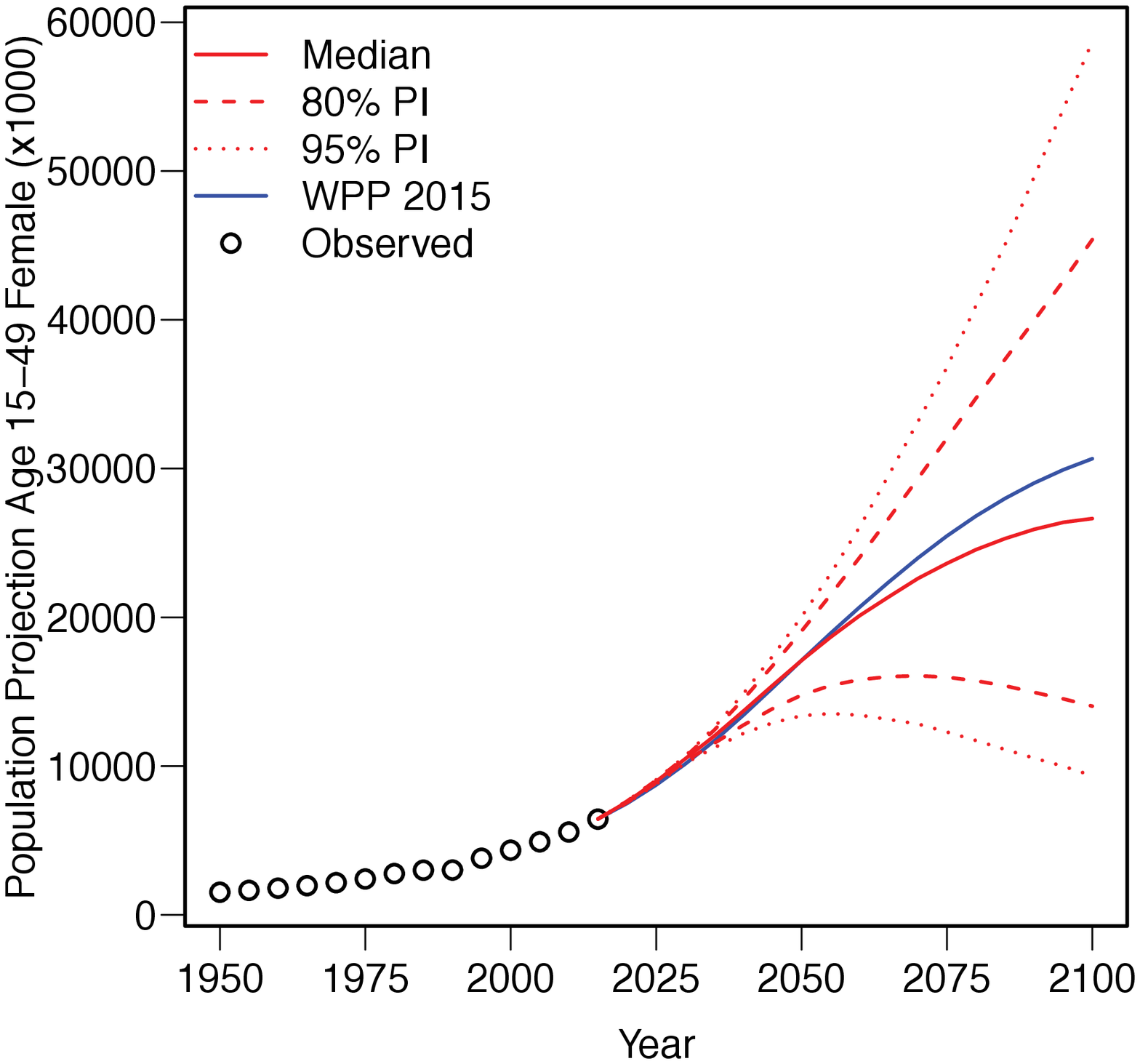}
                \caption{\small Population Projection: female age 15-49}
                \label{fig:ppF1549Mozambique}
        \end{subfigure}
        ~ 
          \begin{subfigure}[b]{0.475\textwidth}
               \includegraphics[width=1\textwidth]{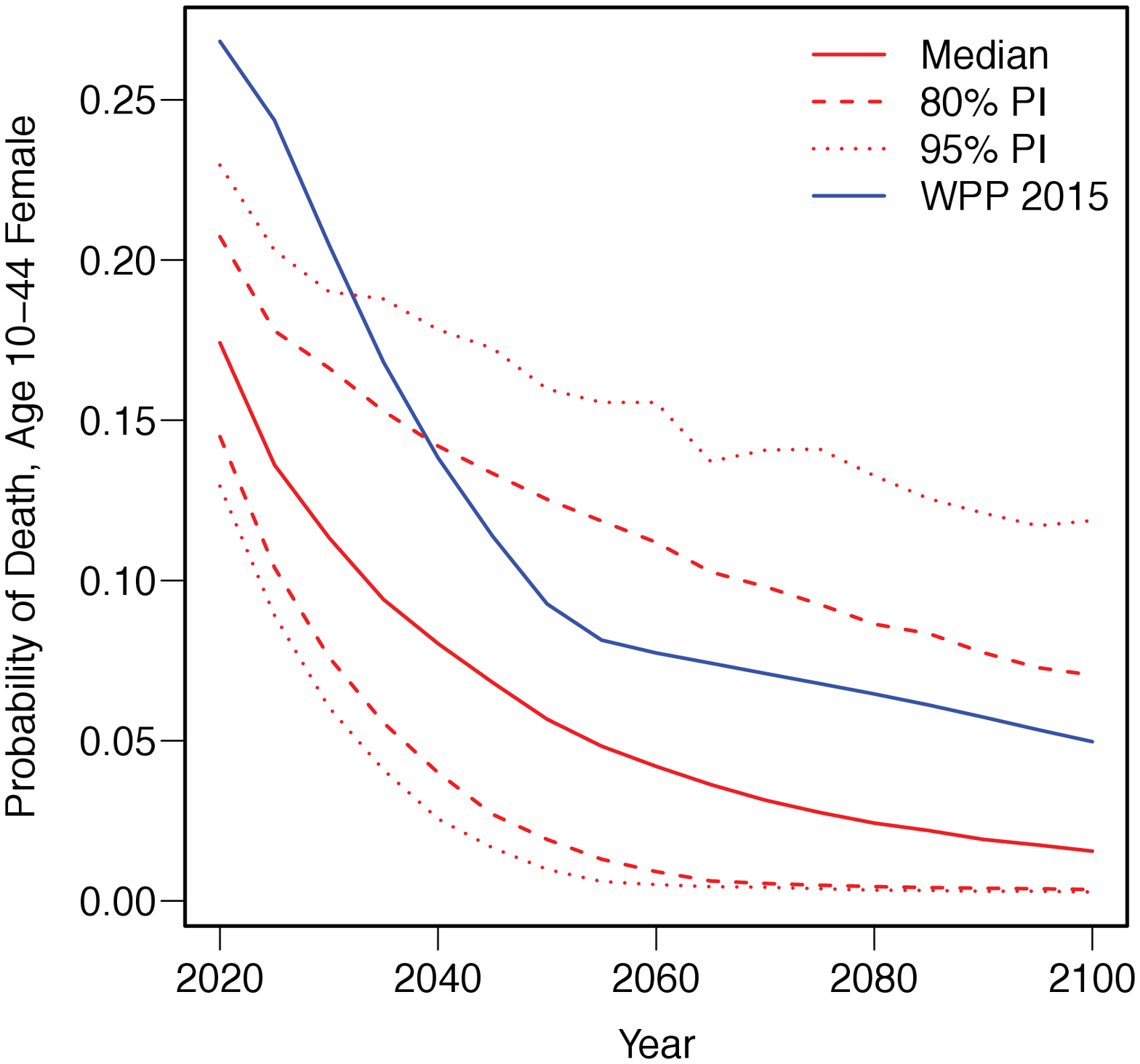}
                \caption{\small Projection of female $_{35}q_{10}$}
                \label{fig:35q10f-Mozambique}
        \end{subfigure}
     
        \caption{Probabilistic population projections for Mozambique 2015-2100. Observed data: black circles; median probabilistic projection: solid red line, 80\% predictive interval: dashed red lines; 95\% predictive interval: dotted red lines; WPP 2015 projection: solid blue line.}
        \label{fig:pppMozambique}
\end{supfigure}

\begin{supfigure}[p]
        \centering
        \begin{subfigure}[b]{.475\textwidth}
               \includegraphics[width=1\textwidth]{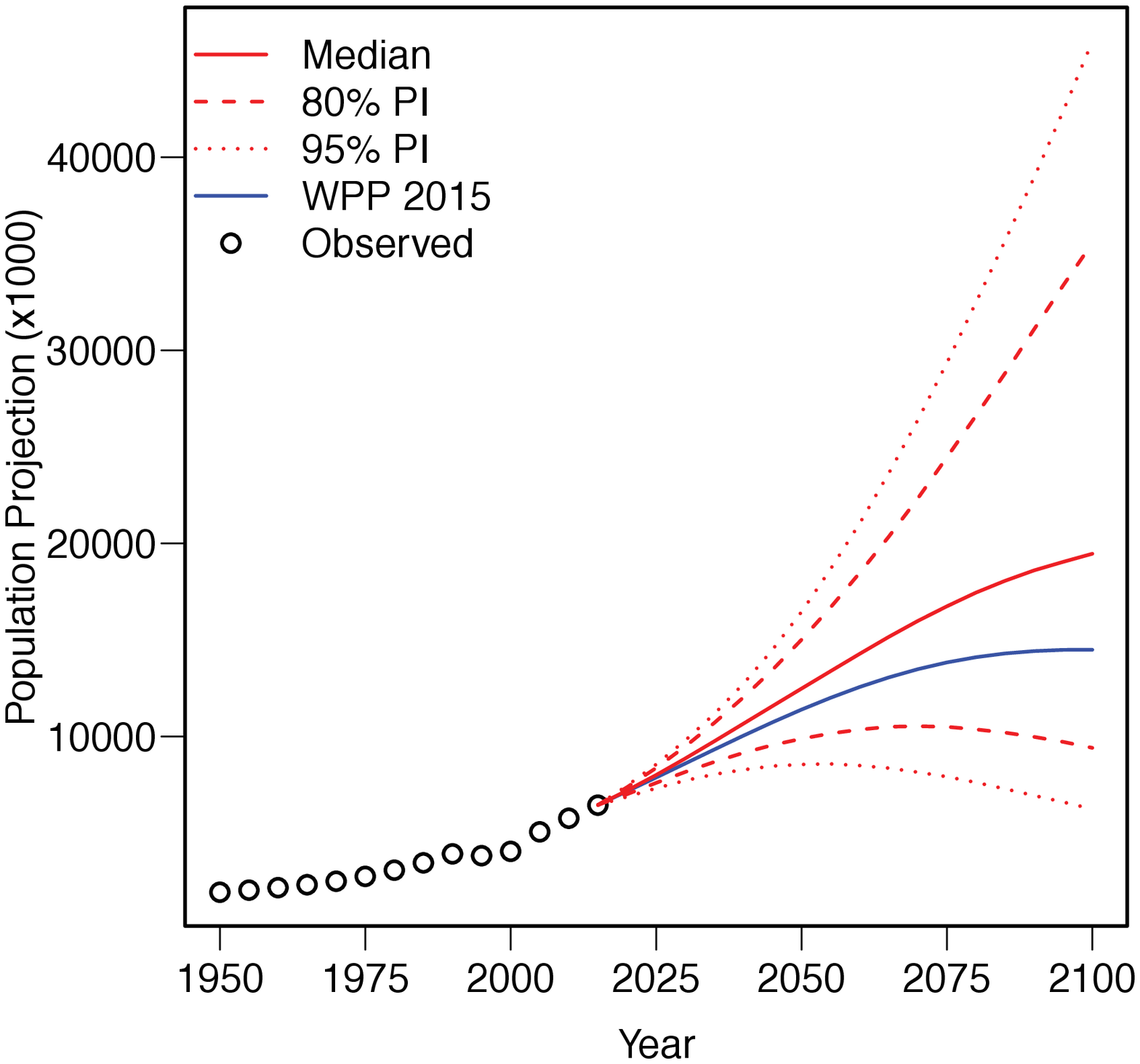}
                \caption{\small Population Projection: total population}
                \label{fig:ppSierraLeone}
        \end{subfigure}%
         ~  
        \begin{subfigure}[b]{0.475\textwidth}
               \includegraphics[width=1\textwidth]{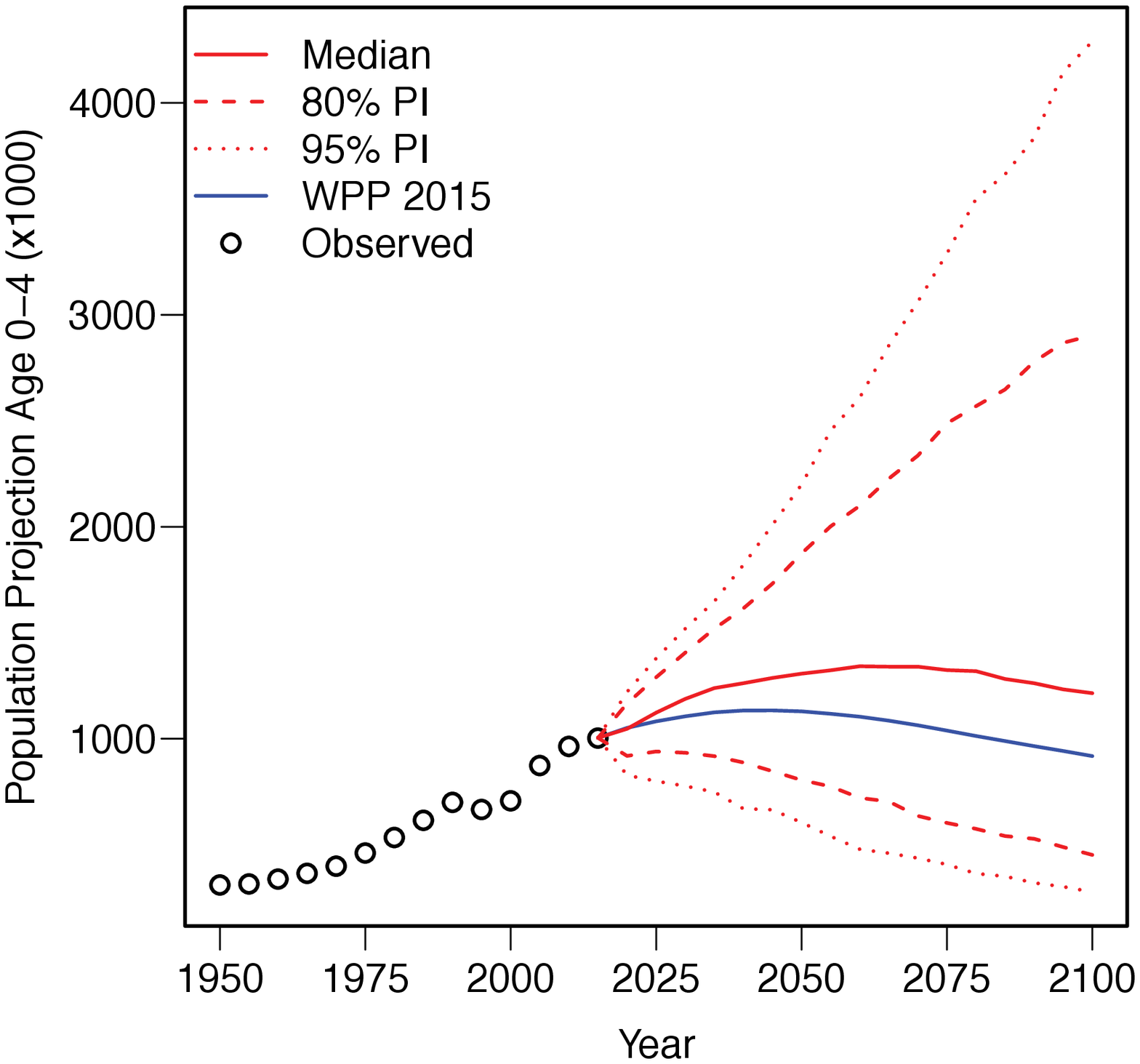}
                \caption{\small Population Projection: age 0-4}
                \label{fig:pp0-4SierraLeone}
        \end{subfigure}
         \\
        \begin{subfigure}[b]{0.475\textwidth}
               \includegraphics[width=1\textwidth]{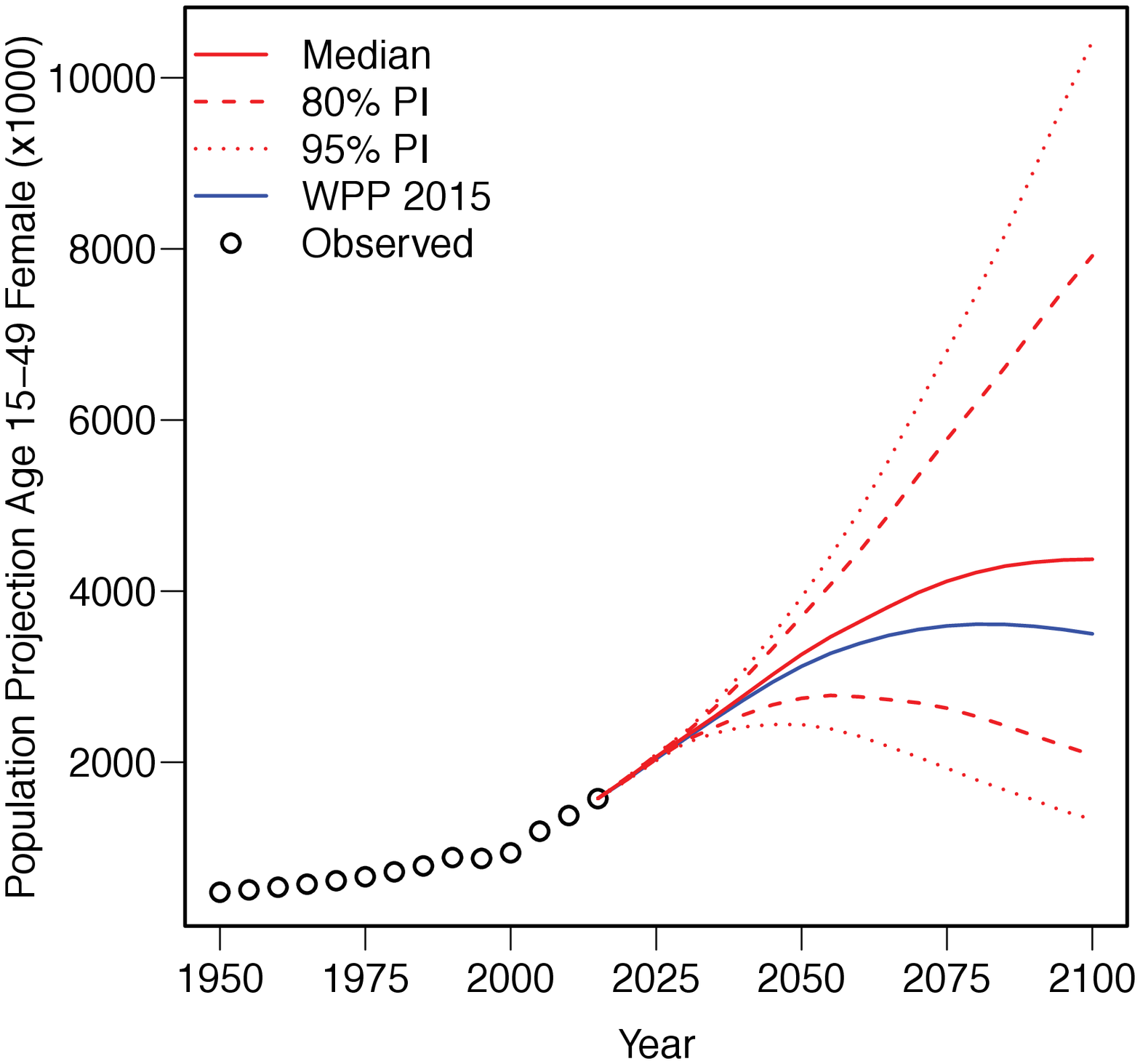}
                \caption{\small Population Projection: female age 15-49}
                \label{fig:ppF1549SierraLeone}
        \end{subfigure}
        ~ 
          \begin{subfigure}[b]{0.475\textwidth}
               \includegraphics[width=1\textwidth]{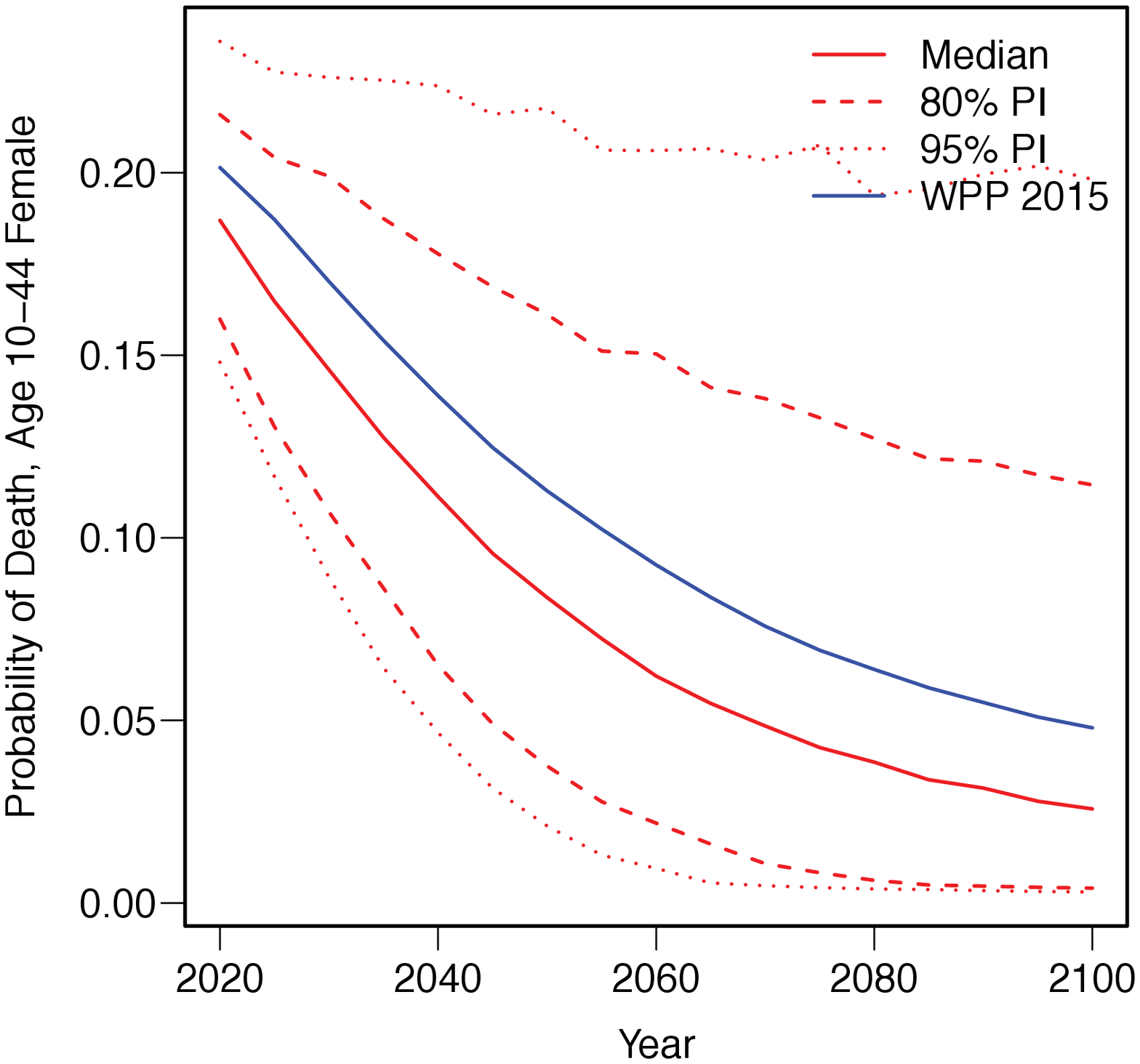}
                \caption{\small Projection of female $_{35}q_{10}$}
                \label{fig:35q10f-SierraLeone}
        \end{subfigure}
     
        \caption{Probabilistic population projections for Sierra Leone 2015-2100. Observed data: black circles; median probabilistic projection: solid red line, 80\% predictive interval: dashed red lines; 95\% predictive interval: dotted red lines; WPP 2015 projection: solid blue line.}
        \label{fig:pppSierraLeone}
\end{supfigure}

\begin{supfigure}[p]
        \centering
        \begin{subfigure}[b]{.375\textwidth}
               \includegraphics[width=1\textwidth]{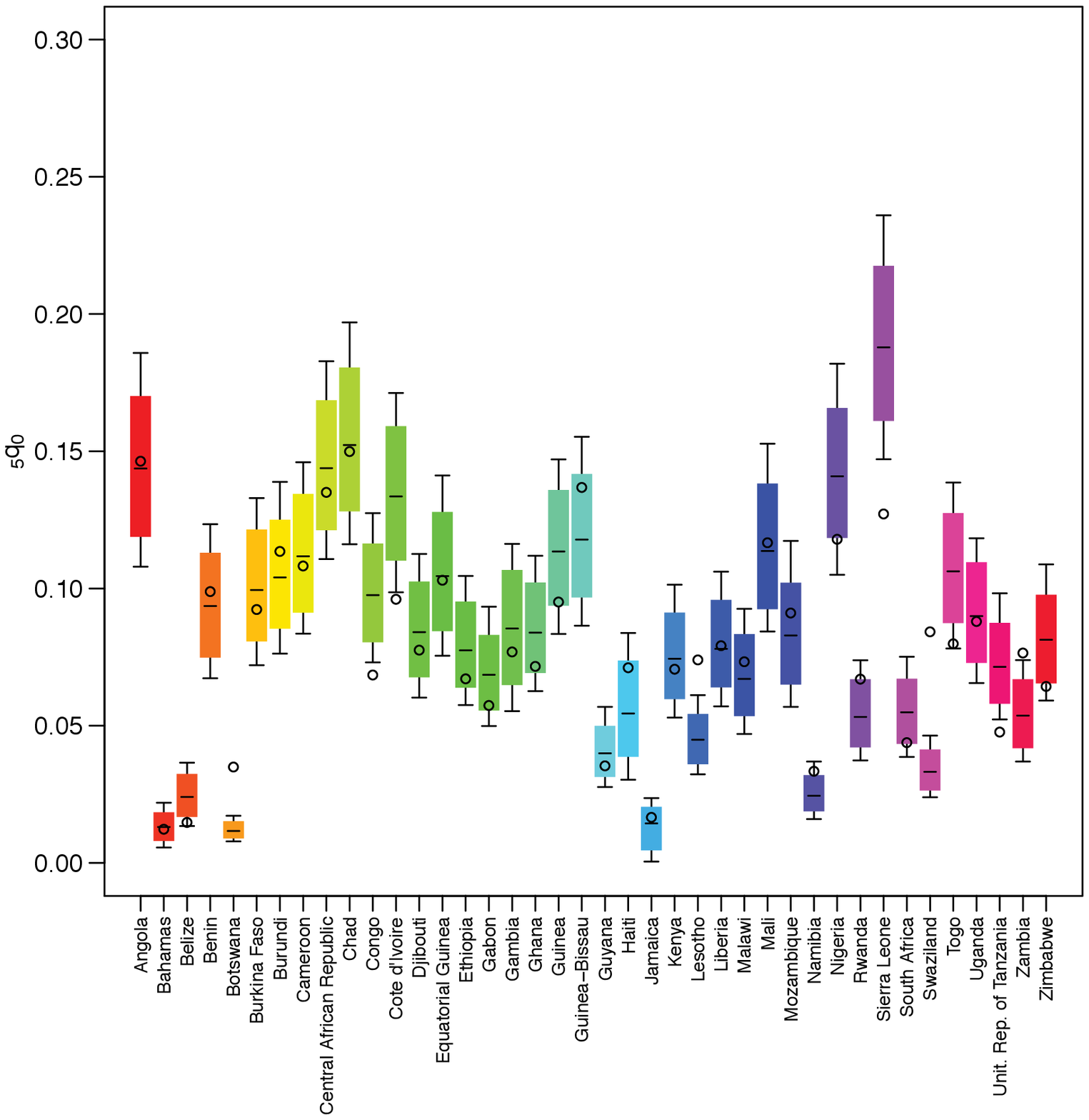}
                \caption{\small $_{5}q_{0}$: female}
        \end{subfigure}%
         ~  
        \begin{subfigure}[b]{0.375\textwidth}
               \includegraphics[width=1\textwidth]{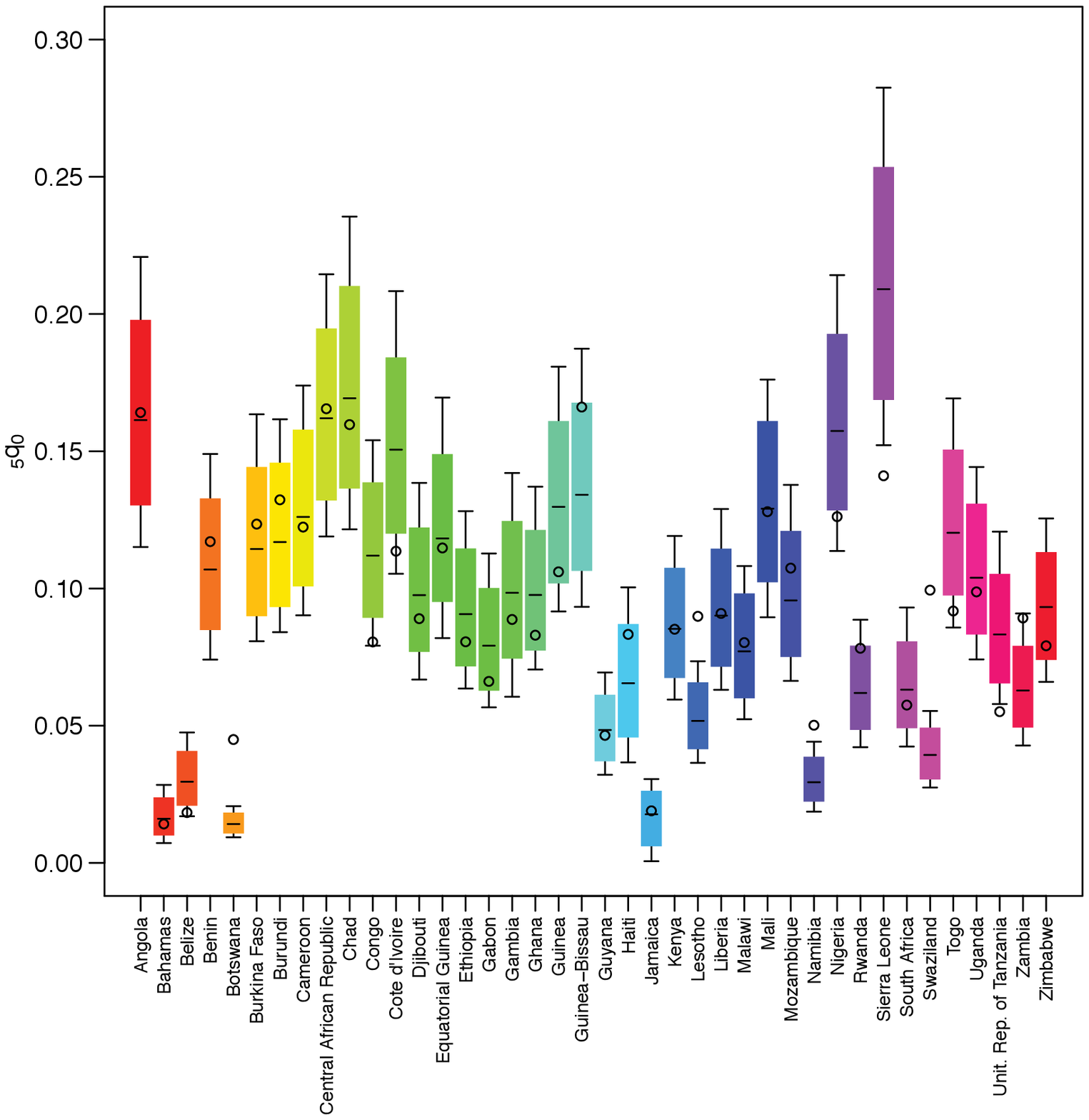}
                 \caption{\small $_{5}q_{0}$: male}
        \end{subfigure}
         \\
        \begin{subfigure}[b]{0.375\textwidth}
               \includegraphics[width=1\textwidth]{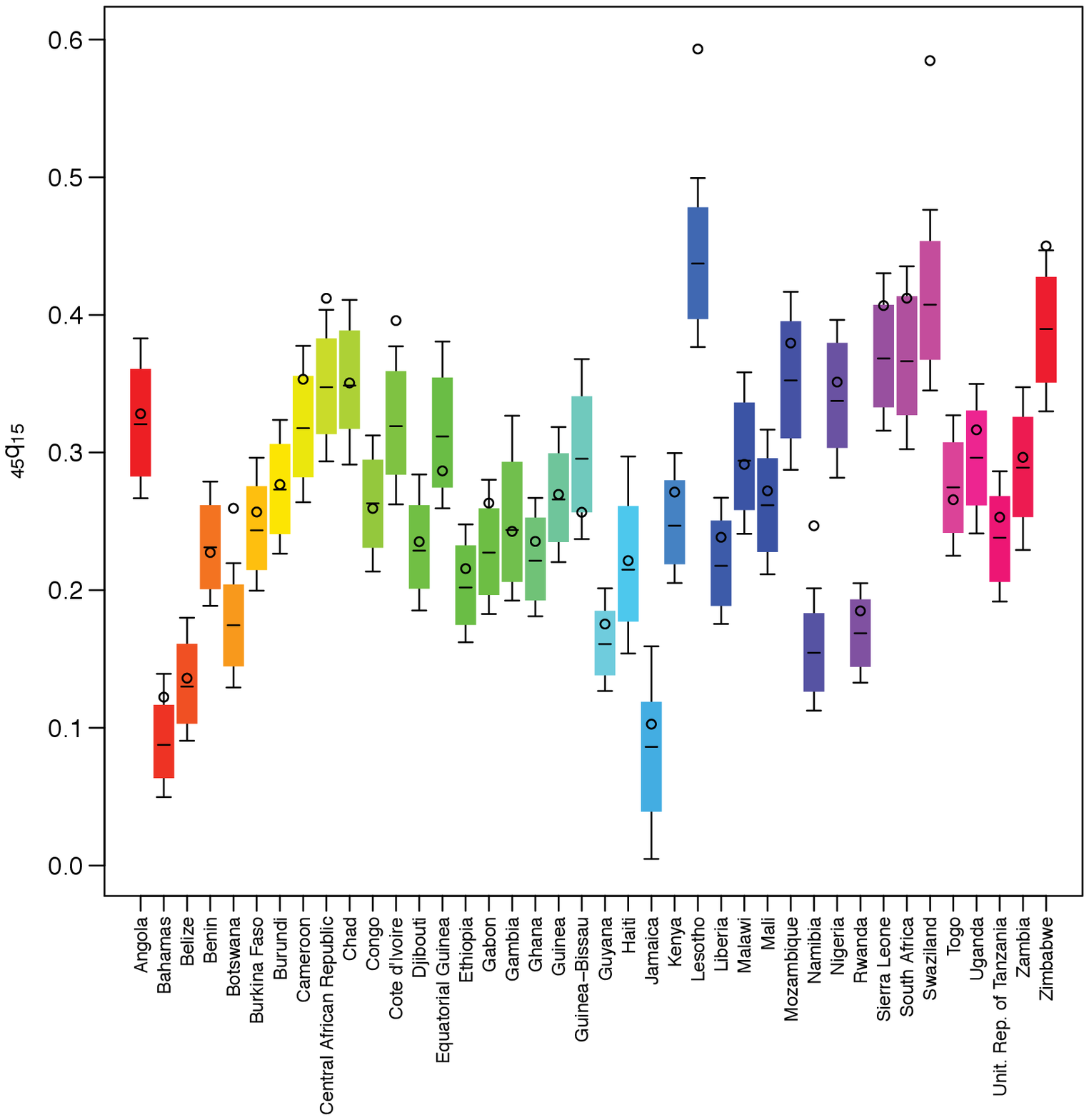}
               \caption{\small $_{45}q_{15}$: female}
        \end{subfigure}
        ~ 
        \begin{subfigure}[b]{0.375\textwidth}
               \includegraphics[width=1\textwidth]{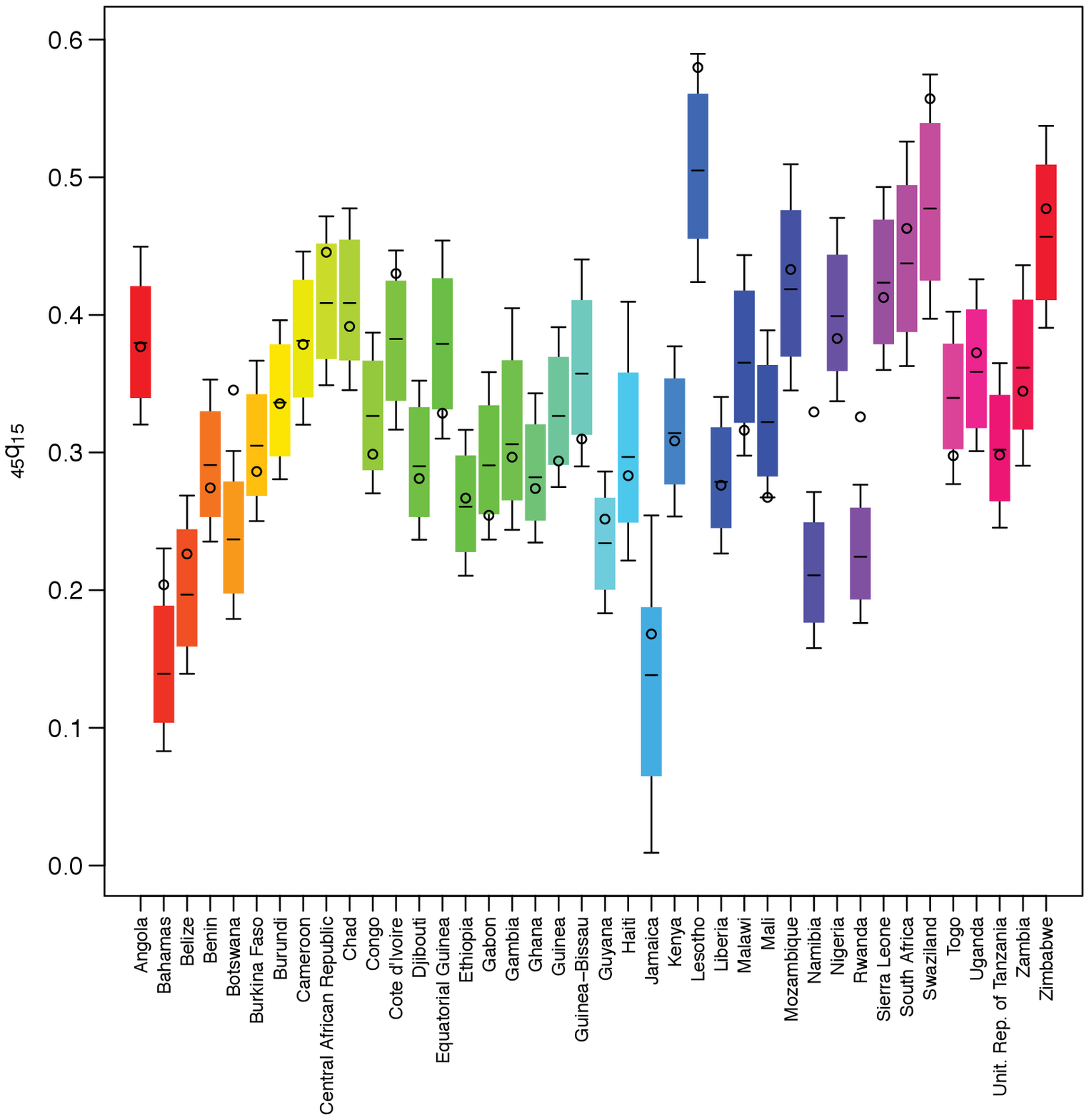}
               \caption{\small $_{45}q_{15}$: male}
        \end{subfigure}
        \\
        \begin{subfigure}[b]{0.375\textwidth}
               \includegraphics[width=1\textwidth]{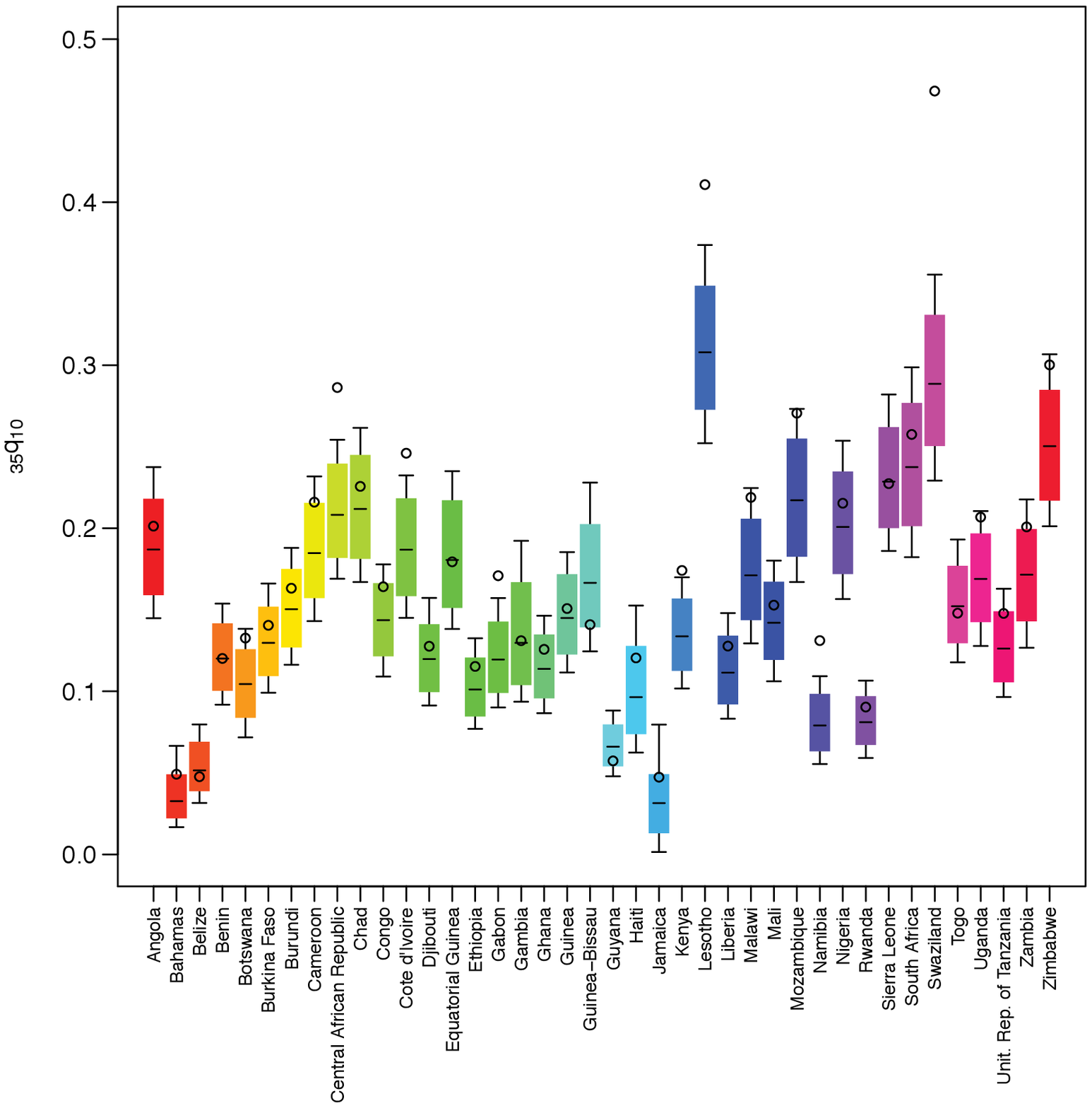}
               \caption{\small $_{35}q_{10}$: female}
        \end{subfigure}
        ~ 
        \begin{subfigure}[b]{0.375\textwidth}
               \includegraphics[width=1\textwidth]{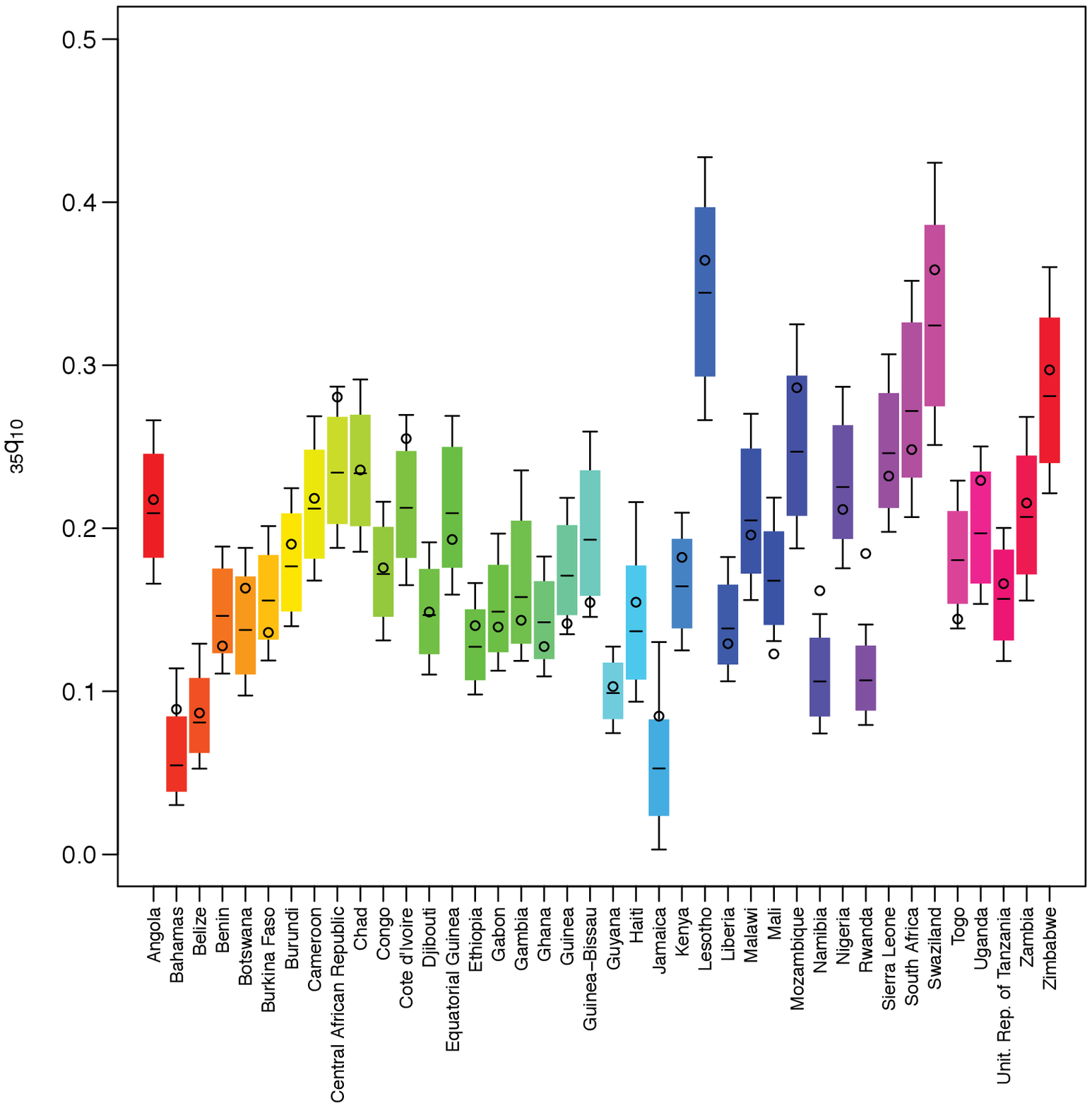}
                \caption{\small $_{35}q_{10}$: male}
        \end{subfigure}
     
        \caption{Distribution of predicted $_{5}q_{0}$, $_{45}q_{15}$, and $_{35}q_{10}$ by country for out-of-sample validation period 2010-2015. In these figures the ends of the ``whiskers'' are the 95\% PI, the ends of the box are the 80\% PI, and the horizontal black line in the center of each box is the median. The WPP 2015 estimate for 2010-2015 is shown with a black circle for comparison.}
        \label{fig:mort-val-dists}
\end{supfigure}

\end{document}